\documentclass[preprint]{aastex}

\usepackage{natbib}
\usepackage{longtable}
\usepackage{lscape}

\begin{document}

\title{Identifying and Quantifying Recurrent Novae Masquerading as Classical Novae}
\author{Ashley Pagnotta\altaffilmark{1}}
\and
\author{Bradley E. Schaefer\altaffilmark{2}}
\altaffiltext{1}{Department of Astrophysics, American Museum of Natural History, New York, NY 10024}
\altaffiltext{2}{Department of Physics and Astronomy, Louisiana State University, Baton Rouge, LA 70803}
\email{pagnotta@amnh.org}

\begin{abstract}
Recurrent novae (RNe) are cataclysmic variables with two or more nova eruptions within a century.  Classical novae (CNe) are similar systems with only one such eruption. Many of the so-called `CNe' are actually RNe for which only one eruption has been discovered. Since RNe are  candidate Type Ia supernova progenitors, it is important to know whether there are enough in our galaxy to provide the supernova rate, and therefore to know how many RNe are masquerading as CNe. To quantify this, we collected all available information on the light curves and spectra of a Galactic, time-limited sample of 237 CNe and the 10 known RNe, as well as exhaustive discovery efficiency records.  We recognize RNe as having (a) outburst amplitude smaller than $14.5-4.5 \times \log(t_3)$, (b) orbital period $>$0.6 days, (c) infrared colors of $J-H>0.7$ mag and $H-K>0.1$ mag, (d) FWHM of  H$\alpha > 2000$ km s$^{-1}$, (e) high excitation lines, such as Fe X or He II near peak, (f) eruption light curves with a plateau, and (g) white dwarf mass greater than $1.2 M_{\odot}$.  Using these criteria, we identify V1721 Aql, DE Cir, CP Cru, KT Eri, V838 Her, V2672 Oph, V4160 Sgr, V4643 Sgr, V4739 Sgr, and V477 Sct as strong RN candidates.  We evaluate the RN fraction amongst the known CNe using three methods to get $24\% \pm 4\%$,  $12\% \pm 3\%$, and  $35\% \pm 3\%$.  With roughly a quarter of the 394 known Galactic novae actually being RNe, there should be approximately a hundred such systems masquerading as CNe.

\end{abstract}


\section{Recurrent Nova Candidates}
\label{sec:intro}
Both classical and recurrent novae (CNe and RNe, respectively) consist of a white dwarf (WD) accreting material from a companion star.  The accreted material accumulates until reaching a critical temperature/pressure at the base of the accreted layer, at which point thermonuclear runaway is triggered and the nova eruption occurs.  The outburst mechanism is identical for both CNe and RNe, but the recurrence timescale varies by multiple orders of magnitude, with RNe seen to erupt at least once per century. The systems classified as CNe have only one {\it discovered} eruption, but more undiscovered eruptions could have occurred within the last century. The truly classical systems do not have any more eruptions on timescales of less than a century. We note that this century-long timescale is empirically based on observations, and is somewhat arbitrary, arising due more to the history of reliably-recorded, large-scale observations (dating back to the 1890s, when the first astronomical plates were made) than to any physical distinction. We anticipate needing to either expand the definition of an RN in the future, as we discover systems with recurrence times just slightly greater than 100 years, or to alter the nomenclature to something such as Fast Recurrence Time Novae, to distinguish systems whose recurrences we have had time to observe from those for which we still wait, a wait time which may be on the order of $10^5$ years. For this paper, we will continue with the current, community-accepted definition that a system with multiple eruptions in fewer than 100 years is called a recurrent nova.

There are two characteristics of RNe that combine to cause their short recurrence time: a high-mass WD (where $M_\mathrm{WD}$ is near the Chandrasekhar mass limit) and a high accretion rate ($\dot{M}$). There is a given trigger mass of accreted material that must be reached for the nova eruption to occur; this trigger mass is smaller for high-mass WDs \citep{yaron2005a} and is more quickly reached when matter is being accumulated at a high rate (high $\dot{M}$), thus the combination of these two factors yields\textemdash and is in fact required for\textemdash the short recurrence times seen in the RNe \citep{sekiguchi1995a,townsley2008a}.

These two factors are also exactly what are needed to cause a WD to explode as a Type Ia supernova (SN Ia), which is why RNe are among the best possibilities for a solution to the long-standing SN Ia progenitor problem.  The reliability of these SNe as standard candles \citep{hamuy1996a} has been key for the measure of the acceleration of the expansion of the universe \citep{riess1998a, perlmutter1999a}.  Questions have been raised, however, about potential evolution in the SN Ia population that could cause peak absolute magnitude variations of up to 0.2 mag and impact the precision cosmology measurements that are made using SNe Ia \citep{dominguez2001a}. An understanding of the dominant progenitor channel, or channels \citep{brandt2010a}, will help answer these questions and reduce the systematic errors which currently dominate the uncertainties in supernova cosmology.  The RN channel is a potential good solution to this key problem, but a critical question is whether there are enough RNe in our Milky Way Galaxy to account for the observed SN Ia rate \citep{branch1995a,della-valle1996a}. A substantial problem with prior work is the lack of consideration for the efficiency of nova discoveries; as we will show, many so-called CNe are actually RNe. This paper provides the first consideration of the numbers and fractions of RNe that are masquerading as CNe, addressing both the RN demographics question and part of the SN Ia progenitor problem. We note that having a population large enough to provide a significant fraction of the SN Ia rate does not in and of itself mean that RNe must be dominant SN Ia progenitors. There are other factors that must be considered, and we discuss these further in Section \ref{sec:implications}, but it is a crucial part of deciding whether RNe are {\it viable} progenitors.

The discovery efficiencies of nova eruptions and RNe are well understood \citep{shafter2002a,pagnotta2009a,schaefer2010b}.  Eruptions over the last century can easily be missed if they happen when the star is too close to the Sun, during the full moon, during any of many intervals when no one was watching, or if the search did not go deep enough even if someone was searching in the right area at the right time.  For CNe, the discovery efficiency is 22\% for novae peaking at $V=6$ mag and 9\% for novae peaking at $V=10$ mag, even in ideal conditions; this rate remains fairly constant from 1890 to 2012. For the ten known RNe, the undirected discovery efficiency for a single eruption varies from 0.6\% to 19\%, with a median of 4\%.  Allowing for multiple eruptions since 1890, an approximately correct situation is a discovery rate of ${\sim}10\%$, so two eruptions are discovered 1\% of the time (and the system is recognized as an RN), and just one eruption is discovered 18\% of the time (so the system is classified as a CN), leaving zero eruptions discovered 81\% of the time. Thus, for every known RN, there must be an order of magnitude more RNe currently masquerading as CNe.  The best place to find new RNe is therefore in the catalog of known CNe.

This paper presents a comprehensive analysis of the RNe hiding in the Galactic nova catalogs.  Section \ref{sec:criteria} systematizes seven criteria for identifying RNe in the nova catalogs on a probabilistic basis. Two of these criteria (concerning the spectra of the systems in outburst; see Sections {\ref{sub:expvel}} and {\ref{sub:excitation}}) are newly developed in this paper.  Section \ref{sec:previous} analyzes the many prior claims of RN candidates, with most such claims now no longer supportable.  Section \ref{sec:candidates} presents an exhaustive compilation of the measured properties (those relevant for our seven RN criteria) of 237 CNe with useful amounts of information (out of 394 currently known novae) as well as the 10 known galactic RNe.  From this information, we identify ten strong RN candidates and 29 systems that are likely RNe.  Section \ref{sec:fraction} presents three independent analyses that measure the fraction of currently cataloged CNe that are actually RNe which have had multiple eruptions within the last century.  Section \ref{sec:searches} presents results from our extensive searches for prior eruptions within archival material from the Harvard and Sonneberg astronomical plate collections.  Part of the original motivation for this paper was to develop a list of confident RN candidates for future exhaustive searches for previous eruptions.  Section \ref{sec:implications} discusses the broad implications of this work, which are dominated by the realization that the total number of RNe in our Milky Way galaxy is ${\sim}160\times$ higher than in prior estimates. 

\section{RN Candidate Criteria}
\label{sec:criteria}
RNe and CNe have substantial overlap in the observed distributions of their properties.  Indeed, this is expected, since many CNe are really RNe.  Nevertheless, a variety of properties are greatly different between the CNe and the RNe.  For example, most RNe have orbital periods longer than 0.6 days, while most CNe have orbital periods shorter than 0.3 days.  Such properties can be used as indicators for recognizing RNe among the CNe.  Due to the overlapping distribution of properties, no one property (other than multiple observed nova eruptions) can be used to definitively identify the CN or RN nature of any system.  We never {\it prove} that a system is an RN by any means other than finding multiple eruptions.  The presence of multiple positive indicators, however, especially if none are contrary, can make a strong case for the RN nature of a system.

We collect here five known RN indicators plus two strong new ones, for a total of seven indicators, described in the first seven subsections, Sections {\ref{sub:ampt3}}-{\ref{sub:wdmass}}. The final subsection, Section {\ref{sub:triplepeaked}}, describes a third new indicator that is not yet well-enough known or understood to be used in our selection criteria, but is interesting enough that it should be included, in hopes of inspiring further study. Importantly, we sketch out the physical bases for the indicators, so that they are much stronger than simple empirical correlations. Table \ref{tab:indicatorfractions} lists each indicator and its efficiency, and each indicator is developed in detail in the following subsections. 

We note that we have used some of these characteristics in a previous paper \mbox{\citep{pagnotta2009a}} to identify V2487 Oph as a good RN candidate. To avoid introducing any bias into the statistics of this paper, for the purpose of the percentages given in Table {\ref{tab:indicatorfractions}}, we do not include V2487 Oph with either sample, and we give it a unique color in the two figures on which it is plotted.

\subsection{Amplitude/$t_3$}
\label{sub:ampt3}
All RNe have a high $\dot{M}$ and about half of them have a red giant companion. Both of these conditions contribute to RNe being more luminous than CNe during quiescence.  Empiricially, RNe peak at approximately the same absolute magnitude as CNe \citep{schaefer2010b}.  Taken together, these two facts mean that RNe have smaller-amplitude outbursts.  Additionally, RNe tend to have shorter-duration outbursts, as measured by $t_3$, which is the amount of time (in days) it takes for the system brightness to decline three magnitudes from peak.  The faster declines are due to the RNe, with their high-mass WDs, having a smaller trigger mass \citep{yaron2005a} and therefore a smaller eruption envelope mass, within which the photosphere will recede faster, causing a faster decline.  Various researchers have previously noted that RNe empirically have small amplitudes and fast declines, but many CNe were either fast or had small amplitude, so the individual criteria were not good indicators for recurrence.  \citet{duerbeck1987a} combined these two indicators to create a single criterion that selected out almost all RNe and only a few CNe.  This criterion can be expressed in a plot of the nova outburst amplitude ($A$) versus decline speed ($t_3$), where the RNe are both fast {\it and} small-amplitude.  With this `Duerbeck plot', the few CNe within the RN region become prime RN candidates.

\begin{figure}
\centering
\epsscale{1.0}
\plotone{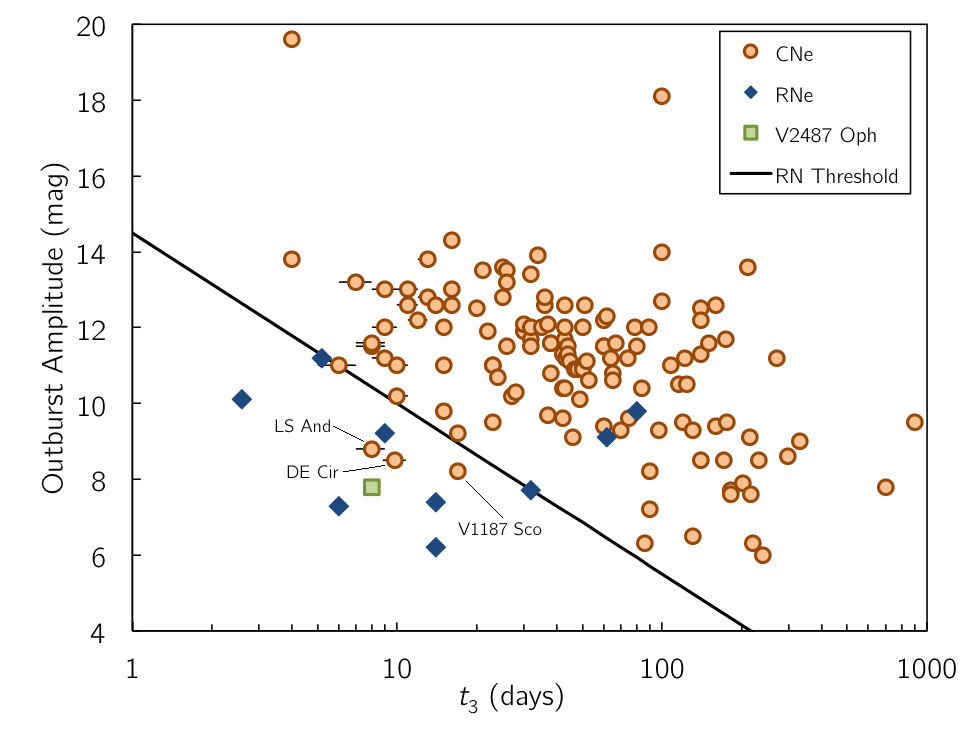}
\caption{This relation, first published in \cite{duerbeck1987a}, plots the amplitude of the nova eruption against the time (in days) to decline by 3 mags from peak, $t_3$; in this plot, the errors are smaller than the symbol size unless otherwise visible. All novae peak at approximately the same absolute magnitude, but RNe have higher average accretion rates and therefore brighter average quiescent magnitudes. This leads to small eruption amplitudes. Additionally, because of the smaller trigger masses required for RNe, the eruptions are shorter and faster than in CNe, so the $t_3$ values are smaller. The RNe (dark blue diamonds) are therefore clustered in the bottom left corner of this plot, with low amplitudes and low $t_3$ values. T Pyx and IM Nor are the notable exceptions, found mixed in with the CNe (brown circles) above the threshold, which is not surprising since they are unusual systems. To quantify the region that \cite{duerbeck1987a} described as ``void of classical novae", we define a threshold line of $A_0=14.5-4.5 \times \log{t_3}$, which is drawn on the plot. 77.8\% of the considered RNe have $A-A_0$ values $<0$, while only three CN systems (2.3\% of our sample) do. Those three systems (LS And, DE Cir, and V1187 Sco) are marked on the figure above. Another six systems (V868 Cen, CP Cru, V4361 Sgr, V697 Sco, V723 Sco, and V477 Sct) have $A-A_0 < 1$, marking them as interesting. V2487 Oph is plotted as a green square to keep it separate from both the RNe and the CNe; see Section {\ref{sec:criteria}} for an explanation of the reasoning behind the decision to keep V2487 Oph separate.}
\label{fig:ampt3}
\end{figure}

We refine the Duerbeck plot in  Figure \ref{fig:ampt3} by adding the RNe which have been identified since 1987, and by using improved measures of the amplitude and $t_3$ \citep{schaefer2010b,strope2010a}.  This allows us to better define the region Duerbeck described as ``void of classical novae".  We define the amplitude of the edge of this region with the relation $A_0 = 14.5-4.5 \times \log(t_3)$.  We can quantify the position of any nova on the Duerbeck plot by its distance from this threshold line, $A-A_0$.  Seven out of nine (77.8\%) of the considered RNe have negative $A-A_0$, whereas only 2.3\% (3 of 131) of the CNe are inside the $A-A_0<0$ region of the Duerbeck plot.   We note that T Pyx and IM Nor are outliers among RNe (with $A-A_0$ values of 2.7 and 3.9 mag, respectively), which is not unexpected due to the unusual nature of the systems \citep{schaefer2010b}.  With this, we recognize that novae with $A-A_0<0$ are very likely RNe, while novae with something like $A-A_0>5$ mag are very likely CNe. 

The difference between 77.8\% and 2.3\% is sufficiently large that this appears significant.  Nevertheless, we should calulate a formal significance for whether this difference is real, and we do this with two methods.  The first method is to calculate the 1-$\sigma$ uncertainties in the fractions from the usual binomial equation.  The fractions meeting our criterion are $77.8\% \pm 13.9\%$ for the RNe and $2.3\% \pm 1.3\%$ for the CNe.  With this, the two fractions differ at the 5.4$\sigma$ level.  The second method is to use a two-sample Kolmogorov-Smirnov (K-S) test \mbox{\citep{press2002a}}, which is a parameterless test to see whether the RNe are different from the CNe for their distribution of $A_0 = 14.5-4.5 \times \log(t_3)$.  The quantity D is the maximum difference between the distribution fraction, for which our criterion is close to this maximum, so D will be close to $0.778-0.023=0.755$ or slightly larger.  The total sample sizes are 9 for the RNe and 131 for the CNe.  If the RN and CN distributions are drawn from the same parent population, then the probability of D being greater than 0.755 is 0.000136.  Thus, the RN and CN distributions are different at the 7400:1 confidence level, and this is adequate to make a confident conclusion.

We note that this dividing line is entirely empirical, based on the location of the ``normal" RNe on the Duerbeck plot, not on any specific model. It was constructed to quantify a region that was previously defined only with a vague description, but its precise location (equation) is inherently uncertain. We are confident in the candidacy of the systems with $A-A_0 <0$, but those on or near the threshold are also potentially good candidates. To quantify these possible (as opposed to probable) candidates, we also note all of the systems with $A-A_0<1$, which is close enough to the threshold to still be interesting, but not so far that it encompasses a large number of almost-certainly CNe.

A substantial problem is that other classes of `novae' can occupy the RN region of the Duerbeck plot.  Most commonly, large-amplitude dwarf novae can be confused with RNe.  Fortunately, most dwarf novae have amplitudes smaller than RNe, and dwarf novae can be recognized uniquely  from their spectral lines during outburst as well as their recurrence timescales of less than a few years.  X-ray novae (e.g., V404 Cyg and V616 Mon [Nova Mon 1917 and 1975]) are another class with potential for confusion.  X-ray novae are caused by accretion disk instabilities in binaries with black holes.  The amplitude is typically 5-8 mag while the decline time is typically 40-120 days \citep{chen1997a}, so many of these systems will appear to satisfy our $A-A_0<0$ criterion.  These systems can be distinguished by their bright {\it hard} X-ray luminosity during outburst as well as by their distinct light curve morphologies \citep{chen1997a}.  Symbiotic novae also have small amplitude outbursts, typically with amplitudes from 2-6 magnitudes \citep{kenyon1986a}.  Symbiotic novae (e.g., RR Tel and PU Vul) are greatly different from novae on symbiotic stars (e.g., RS Oph and T CrB), despite the unfortunate similarity in names.  Symbiotic novae can be distinguished by the presence of a red giant star with heavy stellar winds, resultant long orbital periods (from years to decades), and incredibly long decline times (from years to over a century).  Any well-observed nova event can be easily distinguished from dwarf novae, X-ray novae, and symbiotic novae, but a sparsely observed event can easily be confused between the classes.

Figure \ref{fig:ampt3} also shows 131 CNe, with the input data coming from Table \ref{tab:bigtable}, as described in Section \ref{sec:candidates}.  Three CNe\textemdash LS And, DE Cir, and V1187 Sco\textemdash are definitely located in the RN region, as is V2487 Oph, which is colored differently for clarity.  Therefore the criterion that $A-A_0<0$ is satisfied by only 2.3\% of the CNe.  Six more systems are within 1 mag of the threshold ($0<A-A_0<1$) and therefore also noted as interesting: V868 Cen, CP Cru, V4361 Sgr, V697 Sco, V723 Sco, and V477 Sct.

\subsection{Evolved Companion Stars I: Long Orbital Periods}
\label{sub:evolved1}
Empirically, we know that most (seven out of nine, or 77.8\%) of the considered RNe have evolved companion stars, either sub-giants (e.g. U Sco) or red giants (e.g. RS Oph). Physically, this makes sense because the evolutionary expansion of these companions easily provides (via Roche lobe overflow) the high $\dot{M}$ needed to drive the fast accumulation of accreted material on the surface of the WD \citep{schaefer2010b}. There are two ways we can identify an evolved companion star in nova systems: long orbital period, $P_\mathrm{orb}$, and infrared color excess. 

Roche Lobe geometry tells us that $P_\mathrm{orb} \propto R_\mathrm{donor}^{1.5}$, where $R_\mathrm{donor}$ is the radius of the mass donating star, so a large evolved companion requires  a long $P_\mathrm{orb}$. 77.8\% (seven of nine) of the considered galactic RNe have orbital periods longer than 0.6 days.  This is starkly different from CNe, for which only 21.0\% (13 of 62) are longer than 0.3 days, and only 12.9\% (8 of 62) are longer than 0.6 days.  The difference between 77.8\% and 12.9\% implies that, all else being equal, a $P_\mathrm{orb}>0.6$ day system is likely an RN.  The threshold of 0.6 days is somewhat loose, as a shorter-period system could contain an evolved star driving a high $\dot{M}$.  But the age of our galaxy and the main sequence lifetime of stars combine to put a lower limit of roughly 0.3 days, below which there can be no evolved donor star.  So we can set a formal criterion of $P_\mathrm{orb}>0.6$ days, while acknowledging that systems with orbital periods from 0.3-0.6 days also have a chance of having an evolved companion star.

The difference between 7 out of 9 (for the RNe with known orbital periods) and 8 out of 62 (for the CNe) satisfying $P_\mathrm{orb}>0.6$ days is large.  For binomial statistics, the fractions with $P_\mathrm{orb}>0.6$ days are $77.8\% \pm 13.9\%$ for the RNe and $12.9\% \pm 4.3\%$ for the CNe, with these fractions being different at the 4.5$\sigma$ level.  The K-S test produces a probability of 0.00268 (373:1) which is just past the usual 3$\sigma$ confidence threshold.  So the difference in the orbital period distribution between RNe and CNe is significant.

\subsection{Evolved Companion Stars II: Red Giant Companions}
\label{sub:evolved2}

\begin{figure}
\centering
\epsscale{1.0}
\plotone{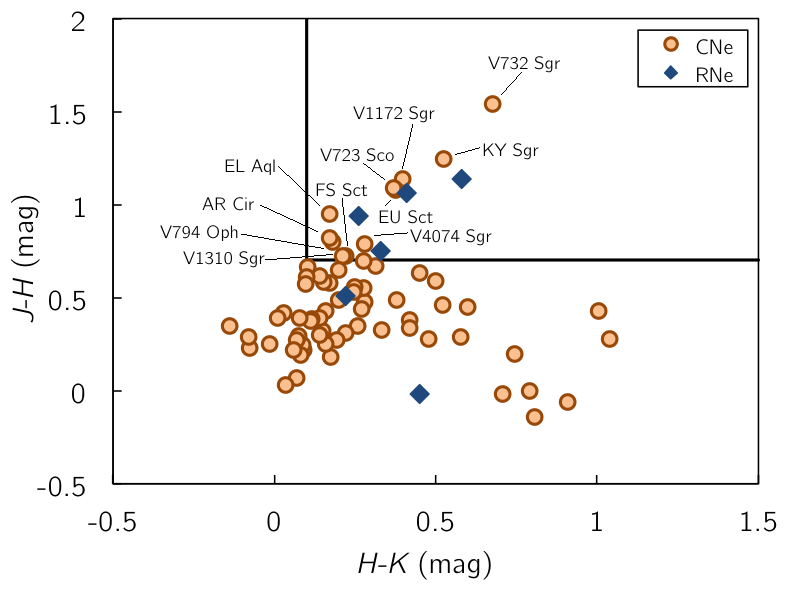}
\caption{This color-color diagram plots two near-IR colors and allows us to identify stars with an infrared excess indicative of a red giant companion. (We note that the average errors on both $J-H$ and $H-K$ are smaller than the symbol size for each point.) Our threshold, which is based on Figure 1 of \mbox{\citet{harrison1992a}}, is $J-H>0.7$ and $H-K>0.1$. This identifies 11 CNe as likely having red giant companions: EL Aql, AR Cir, V794 Oph, KY Sgr, V732 Sgr, V1172 Sgr, V1310 Sgr, V4074 Sgr, V723 Sco, EU Sct, and FS Sct. Systems with red giant companions are potential RNe since the evolutionary expansion of the red giant can easily drive the high accretion rate needed for a short recurrence time, however it is not a given that these systems are recurrent, as RNe also require a high mass WD. Nevertheless, these are interesting systems.}
\label{fig:ircolors}
\end{figure}

The presence of an evolved companion can also be inferred by looking at the infrared colors of the system in quiescence. An evolved companion, particularly a red giant, locates the system in a distinct area on a $J-H$ vs. $H-K$ color-color diagram, as seen in Figure \ref{fig:ircolors}.  The region with evolved companion stars has imperfectly defined edges, but we take it to be $J-H>0.7$ mag and $H-K>0.1$ mag; this threshold is marked in Figure \ref{fig:ircolors}. The cut-offs are loosely based on Figure 1 of \mbox{\citet{harrison1992a}} and are designed to encompass all possible red giants, but are admittedly not perfect due to model dependence. Four of the known RNe have red giant companions and occupy this region of the infrared color-color plot.  Only 15.3\% (11 of 72) of the CNe have colors which indicate the presence of a red giant.  The contrast between 44.4\% (4 of 9) and 15.3\% is not high, so we consider the infrared colors to be only a weak indicator. The fractions with red giants are $44.4\% \pm 16.6\%$ for the RNe and $15.3\% \pm 4.2\%$ for the CNe, with these fractions being different at the 1.7$\sigma$ level.  The formal K-S test gives a 50\% probability that such a difference in fraction could occur by chance sampling of the same parent distributions.  While the detection of a red giant companion is an indicator of low significance, many research groups (see references in Section 3) have adopted this criterion, so it is important to discuss and include it in this section.

We note that fewer than half (only four of nine considered) of the known RNe have red giant companions, so this is certainly not a requirement that must be met in order to consider a system recurrent. It is the presence of any type of {\it evolved} companion that drives the high $\dot{M}$, but at the same time, CNe can have evolved companions as well.  RNe must also have a WD mass near the Chandrasekhar limit, while a system with a lower mass WD will necessarily have a long recurrence timescale \citep{yaron2005a}.  The presence of an evolved companion will substantially increase the expectation that the system might be an RN, but this property alone cannot provide any high confidence. 

\subsection{High Expansion Velocity}
\label{sub:expvel}
The expansion velocity of the ejected material in nova eruptions is measured using widths of emission lines.  The spectra and line widths in the literature cover a wide variety of spectral ranges and elements, which vary in complex manners, but the one nearly universally covered wavelength is that of the H$\alpha$ line at 6563\AA.  The width of the H$\alpha$ line is measured and reported as the FWHM or the FWZI, but the latter quantity is difficult to measure consistently with accuracy.  Therefore, we chose to use the full-width at half-maximum (FWHM) of the H$\alpha$ line as a standard measure because this line is ubiquitous in nova spectra and the FWHM of the line is commonly reported. 

RNe should systematically have high FWHM. RNe are distinguished from CNe by having a WD near the Chandrasekhar mass, and such WDs must have high mass and a small radius (when compared to the WDs in CNe).  This necessary condition will lead to the WD having a high escape velocity, $v_{esc}=\sqrt{{2GM_\mathrm{WD}} r_\mathrm{WD}^{-1}}$, much higher than for any CN. For detailed numbers, the WD escape velocity is 4300, 6800, 9000, and 12700 km s$^{-1}$ for WDs with masses 0.6, 1.0, 1.2, and $1.35 M_{\odot}$, respectively. Roughly, the RNe should have WD escape velocities that are double or triple those of CNe.  In general, for explosive events, the typical ejection velocity is comparable to the escape velocity.  Hence, RNe should have double or triple the ejection velocities of CNe, and the RN H$\alpha$ FWHMs should be double or triple those of CNe. \citet{kato2003a} have made detailed models of the velocity outflows in RNe, which can achieve the high observed velocities. They point out that the ejection velocity depends sensitively on the WD mass, but this is the first time expansion velocity has been used explicitly to identify new RN candidates.

\begin{figure}
\centering
\epsscale{1.0}
\plotone{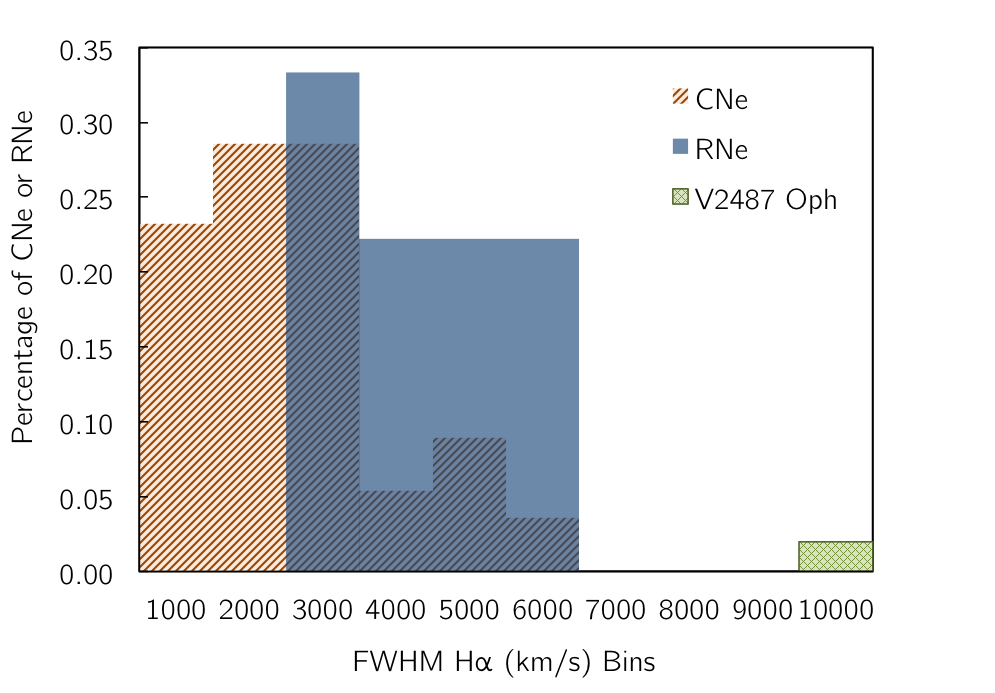}
\caption{This overlapping histogram shows the distribution of the expansion velocity of nova eruptions, as measured by the full-width half-maximum (FWHM) of the H$\alpha$ line in km s$^{-1}$ as close to peak as possible; the bin value labels along the $x$-axis are the (inclusive) maximum FWHM for that bin, so the first bin contains all systems with 0 km s$^{-1} \le $ FWHM $\le$ 1000 km s$^{-1}$, the second bin contains all systems with 1000 km s$^{-1} < $ FWHM $\le$ 2000 km s$^{-1}$, and so forth. The CNe (brown diagonal hashed region) have on average much lower expansion velocities, due to their less massive WDs. The RNe (light blue solid region) have much higher expansion velocities because of the presence of a high-mass WD in the system. The populations are not completely distinct, as can be seen by the overlap in the distributions in this plot. Part of this is due to the natural continuum of nova eruption characteristics, and part is due to the fact that there are many RNe miscategorized as CNe. We consider any CN with FWHM H$\alpha >$2000 km s$^{-1}$ to be a possible RN, and any CN with FWHM H$\alpha >$ 3500 km s$^{-1}$ to have a high probability of being recurrent. As it has been throughout this paper, V2487 Oph is treated separately from both the RN and CN samples; to that effect, it is depicted with green cross-hatching on the plot, at its observed expansion velocity of 10000 km s$^{-1}$, at the percentage amount it would be if it were grouped in with the CNe, which is a somewhat arbitrary y-value, chosen mostly to give a low height to that section of the histogram to depict the fact that there is only one object in our time-limited sample out there. See Section {\ref{sec:criteria}} for more on the reasoning behind the decision to keep V2487 Oph separate.}
\label{fig:expvel}
\end{figure}

A histogram showing the distribution of FWHMs of H$\alpha$ for our sample can be seen in Figure \ref{fig:expvel}. Every known RN has a FWHM of H$\alpha$ greater than 2000 km s$^{-1}$, and the median velocity is 3930 km s$^{-1}$.  The highest velocity is 10,000 km s$^{-1}$, for V2487 Oph; excluding V2487 Oph (since we previously used this characteristic to identify it as a strong RN candidate), the next highest RN escape velocity is 5700 km s$^{-1}$ for U Sco.  This is in contrast to the CNe, where only 52.6\% 30 of 57 meet or exceed 2000 km s$^{-1}$.  With this, we take a nova that has FWHM of H$\alpha$ $<$ 2000 km s$^{-1}$ to be a likely CN. We take 3500 km s$^{-1}$ (with 66.7\% [6 of 9] of the RNe and 14.4\% [8 of 57] of the CNe) to be a threshold above which the systems have a high probability of being recurrent.  The binomial uncertainties on these fractions show them to be different at the 7.1$\sigma$ and 3.1$\sigma$ level  for criteria of $>$2000 and $>$3500 km s$^{-1}$ respectively.  The K-S test gives probabilities of 0.061 and 0.036 for the RN and CN distributions of FWHM both arising from the same parent population, for velocities of $>$2000 and $>$3500 km s$^{-1}$ respectively.  The difference between CNe and RNe looks to be substantial in Figure {\ref{fig:expvel}}, and our statistical tests show that the difference is likely real.

On this basis, from our time-limited sample of 237 novae, DE Cir, V693 CrA, V838 Her, V4160 Sgr, V4643 Sgr, V4739 Sgr, and  V1142 Sco are good RN candidates.  From later novae that are not part of this sample, V1721 Aql, V2491 Cyg, KT Eri, and V2672 Oph are also strong RN candidates.

\subsection{High Excitation Lines}
\label{sub:excitation}
RNe also display unusually high excitation lines in their early outburst spectra. Again this is due to the high-mass WD \citep{starrfield2012a}, which sits inside a deep potential energy well. Large amounts of energy are needed to propel the ejected material out of this well, and this is reflected in the spectra, but has never been used as a criterion for RN candidates systems until now.

We use two primary indications of high-excitation: the He II and Fe X emission lines.  Sometimes the highest excitation conditions produce Fe XI and Fe XIV lines, as well as Ne V lines.  These emission lines are sought in the spectra from near-peak and soon thereafter.  Spectra taken too long after the peak will often pick up the He II emission line that is normal for the nebular phase; we must avoid including these since the source of this excitation is not related to the effect we are seeking here to identify possible RNe.

For the nine RNe considered for these statistics, five of eight show high-excitation iron lines (ranging from Fe X to Fe XIV), and all nine show He II lines. The presence of Fe X (or above) or He II lines is thus a strong criterion for candidate RNe. We can say with reasonable confidence that any nova that does not show either He II or Fe X, when both were sought, is likely a CN. For CNe, \citet{williams1993a} classifies the majority as Fe II systems, the quintessential case for low excitation. In our sample, we have 50 systems for which He II was sought; of those, 13 (26.0\%) show He II lines near optical peak. There are 64 systems for which iron lines were reported; of those, only four (6.3\%) show Fe X or higher. There is one system, CP Pup, which shows both. Although all of the known RNe show He II in their early spectra, we do not consider a lack of He II to preclude a system from being an RN. If that same system does have high iron lines, it is still reasonable to mark it as a good RN candidate because the iron is an indicator of a powerful eruption, and it is possible to imagine situations in which the He II went unobserved. In total, there are 77 systems for which at least one spectral indicator was sought, and 16 of those showed early He II and/or Fe X, for a total of 20.8\%. Following the same procedure, all nine of the considered RNe show early He II and/or Fe X, so in Table 1 the 20.8\% CN figure is compared to 100\% for RNe. The binomial probability gives the formal uncertainties in the fractions to be $100\% \pm 0\%$ (although a realistic uncertainty would be larger than $\pm 0 \%$ as formally given by the usual binomial result) for the RNe and $20.8\% \pm 4.6\%$ for the CNe.  The K-S test shows this criterion to distinguish RNe and CNe at the 12300:1 level.

\subsection{Light Curve Plateaus}
\citet{hachisu2008a} have pointed to eruption light curve plateaus as RN markers.  The fast emergence of the supersoft X-ray source irradiates the re-formed accretion disk and companion star, providing reprocessed optical light that\textemdash for a time\textemdash dominates over the fast fading shell light.  The supersoft X-ray luminosity does not vary greatly, until beginning a fairly fast turn off, resulting in a plateau in the optical light curve.

The \citet{strope2010a} analysis defines seven light curve classes based on the light curve shapes.  The most common light curve classes are S (smoothly declining), D (dust dips), and J (large random jitters from before the peak until the transition phase); other classes include O (periodic oscillations around the transition phase), F (flat-topped), and C (a slow-rising fast-fall cusp around the transition phase).  Of importance here, the P class light curves are characterized by a plateau around the transition phase, with this plateau being an interruption of the otherwise-smooth decline from peak, at which point the light curve goes nearly flat for a measurable amount of time, then abruptly returns to its steep decline, often at a faster rate than before the plateau phase.  \citet{strope2010a} calculate that  only 16.9\% (13 of 77) of CNe have plateaus.

The massive collection and analysis of virtually all extant photometry of RNe \citep{schaefer2010b} provides a quantitative and definitive measure of the fraction of RNe that have plateaus in their light curve. \citet{schaefer2010b} and \citet{strope2010a} found that all RN light curves were certainly either S or P class, with 55.6\% to 88.9\% of them being P class, for our nine considered RNe.  The uncertainty in the plateau fraction for RNe is due to poor late-time light curves for three of the systems.  Only one RN (T CrB) was certainly an S class event.

Thus 55.6-88.9\% of RNe are P class, while 16.9\% of CNe are P class.  These differences are significant at the 8.8$\sigma$ level from the binomial uncertainties and at the 2100:1 level from the K-S analysis. This provides good empirical support for the model predictions of \citet{hachisu2008a}.  Hence, a plateau provides a reasonable indicator of a RN, while the lack of a plateau is a reasonable indicator of a CN.  The fractions indicate, however, that this is not a strong case by itself.  A substantially stronger case can be made that all D and J class novae are low energy events \citep{strope2010a} and therefore not RNe.

\subsection{White Dwarf Mass}
\label{sub:wdmass}
One requirement for RNe is that their WD must be near the Chandrasekhar mass.  There is no firm threshold, but grids of nova models \citep{yaron2005a} suggest a lower limit of roughly $1.20{-}1.25 M_{\odot}$.  WD masses have been measured in some nova systems, but mass measurements are notoriously difficult and have large uncertainties. Nevertheless, a high-mass WD is necessary to produce an RN and a low-mass WD appears to preclude an RN, so we can say that a system is a likely RN if $M_\mathrm{WD} \ge 1.2 M_{\odot}$, while a system is likely to be a CN if $M_\mathrm{WD}<1.2 M_{\odot}$.   Presumably, all RNe pass this criterion. This presumption is based largely on theoretical models, and backed up by only a limited number of observations of some RN WD masses, but we are confident that the presumption is robust.  I. Hachisu and M. Kato have a series of papers 
\citep{hachisu2004a,hachisu2005a,hachisu2006a,hachisu2007a,hachisu2008c,hachisu2008a,hachisu2009a,hachisu2010a,kato2003b,kato2007a,kato2009a,kato2011a} wherein they model 34 CN light curves and derive $M_\mathrm{WD}$.  Of the systems listed in Table \ref{tab:bigtable}, 7 of 27 have $M_\mathrm{WD} \ge 1.2 M_\odot$, for a fraction of 25.9\%, and two newer systems have $M_\mathrm{WD} \ge 1.2 M_\odot$ as well. These nine systems are: V693 CrA, V2491 Cyg, V838 Her, V445 Pup, V598 Pup, V5115 Sgr, V1188 Sco, V477 Sct, and V382 Vel. These systems may not necessarily be RNe, as their accretion rate might not be high enough to allow for a short recurrence time scale, but they are certainly still interesting.

\subsection{Triple-Peaked Emission Lines}
\label{sub:triplepeaked}
A newly-recognized, uncommon characteristic that possibly indicates an RN is a triple-peaked structure in the outburst emission lines, most often observed in the Balmer series. This unusual line profile has been identified in three known RNe (U Sco, Nova LMC 2009, and YY Dor, another LMC RN) and a number of suspected RNe, including DE Cir, KT Eri, V2672 Oph, Nova LMC 2012, V5591 Sgr, and HV Cet (F. Walter 2010-2014, private communication), V1721 Aql \citep{helton2008b}, and V838 Her \citep{iijima2010a}.  When such triple-peaked features are seen in supernovae, they are attributed to expanding toruses of gas.  However, \citet{walter2011a} make a case that the triple peaks in RNe come not from an expanding torus, but rather from an accretion disk, which has somehow re-formed very soon after the peak of the eruption.  Currently, there is no complete theoretical explanation for these triple-peaked emission lines.  On the empirical side, only one of the ten known galactic RNe (U Sco) has the triple peaked features, for a rate of 10.0\%, even though most have adequate spectral coverage, so perhaps the low fraction is due to inclination effects.  For CNe, these unusual features have not been reported for many systems, although old spectra of V603 Aql in outburst are described as being triple-peaked \citep{payne-gaposchkin1964a}.  With these numbers and limited coverage, we cannot make this into a standalone RN indicator, and therefore do not include it as one of our seven, or list it as a separate column in Table {\ref{tab:indicatorfractions}}, but such triple peaks apparently increase the likelihood of a system being an RN if they are observed, so we note their presence when they have been observed.

\section{Previously-Proposed RN Candidates}
\label{sec:previous}
Many CNe have been identified as likely RNe in the past, for a variety of reasons. Most are identified because of one particular characteristic, such as infrared colors in quiescence \citep{harrison1992a,weight1994a,hoard2002a}, eruption light curves indicating high-mass WDs \citep{hachisu2002a}, the existence of pre-existing dust possibly left over from previous nova eruptions \citep{kawabata2000a}, or the speed of decline of the outburst light curve \citep{shears2007a}. This is a reasonable method of constructing an initial list of possible RNe, but fails to create a rigorous set of candidates because there are many CNe which display one indicator of a short recurrence time. Considering only one characteristic produces many false positives. \citet{duerbeck1987a} attempted to address this issue by combining two characteristics, outburst amplitude and duration (Section \ref{sub:ampt3}), to identify good RN candidates. We expand this by considering a total of seven characteristics, each of which indicate a short recurrence time. This allows us to focus on systems with multiple strong indicators and therefore maximize our chances of discovering new RNe. 

We are also able to take a broader look at the previously-published candidates to better determine their current status.  Various of these suggestions are weak and should now be forgotten, while others are strong, and we should recognize the original proposers. We discuss each of these systems, listed roughly in GCVS order, in the following subsections.

\subsection{LS And, V794 Oph, V909 Sgr, MU Ser}
\label{sub:lsand}
\citet{duerbeck1988a} points to four CNe as likely RNe because of their positions in the amplitude/$t_3$ diagram.  Unfortunately, nothing is known about LS And other than its position on the Duerbeck plot.  While this points strongly to LS And being an RN, the lack of any other information means that the confidence in this conclusion is not high.  For example, with no additional information, \citet{rosenbush1999a} identified LS And as an X-ray nova.  V794 Oph lies outside our RN regions in both Figures \ref{fig:ampt3} and \ref{fig:ircolors}, and with its 220 day $t_3$, it is almost certainly not recurrent.  V909 Sgr has little known about it other than its position on the Duerbeck plot, but with $A-A_0=2.5$ mag, the case for it being an RN is weak.  For MU Ser, the \citet{duerbeck1988a} amplitude is very uncertain, as the system counterpart is not definitively identified, so it is likely even farther from the threshold than the $A-A_0=0.7$ mag calculated using Duerbeck's values. In all, LS And is weakly likely to be an RN, while the other three are most likely CNe.

\subsection{V368 Aql, V604 Aql, V841 Aql, AR Cir, BT Mon, GK Per, V794 Oph, V1172 Sgr, V3645 Sgr, V723 Sco, EU Sct, FS Sct}
\label{sub:ircolors}
Various groups have used infrared colors as a proxy for a post-main-sequence companion, and hence a likely RN. On this basis, the following candidate RNe have been identified:  V604 Aql, V841 Aql, AR Cir, GK Per, and EU Sct by \citet{szkody1994a}; V3645 Sgr and EU Sct by \citet{weight1994a}; V723 Sco by \citet{harrison1996a}; V1172 Sgr by \citet{hoard2002a}; V368 Aql, BT Mon, V794 Oph, V3645 Sgr, and FS Sct by \citet{surina2011a}. For many of these stars, the presence of an evolved companion might well indicate the recurrent nature of the systems. Nevertheless, there are a variety of problems with some of these candidates. BT Mon and GK Per will be been discussed individually in detail below (Sections \ref{sub:btmon} and \ref{sub:gkper}, respectively) and are certainly not RNe. AR Cir is in fact a symbiotic nova, distinct from a standard classical nova in a symbiotic system \citep{harrison1996a}. V3645 Sgr may also be a symbiotic nova, based on its small eruption amplitude ($\Delta m = 5.4$ mag; \citealp{downes2001a}). EU Sct, V604 Aql, V794 Oph, and V368 Aql fail the amplitude/$t_3$ criterion (Section \ref{sub:ampt3}). Additionally, EU Sct and FS Sct have no high excitation lines (Section \ref{sub:excitation}), so we judge these to be weak cases. Two proposed candidates with post-main-sequence secondary stars, however, present hopeful cases: V1172 Sgr has an H$\alpha$ FWHM of 2000-3000 km s$^{-1}$ \citep{stromgren1951a} as well as IR colors that indicate a giant companion, and V723 Sco has a fast decline ($t_3 = 17$ days; \citealp{shafter1997a}) combined with IR colors that indicate an evolved companion, possibly a giant.

\subsection{V1721 Aql}
\label{sub:v1721aql}
V1721 Aql (Nova Aql 2008) is a poorly observed nova that was never seen brighter than $V=14$ mag, due to high extinction ($11.6 \pm 0.2$ mag) in the $V$ band. \citet{hounsell2011a} describe the various features that indicate an RN. The light curve fades very quickly ($t_3=10$ days). The FWHM of H$\alpha$ is 6450 km s$^{-1}$, and the outburst spectrum shows broad, triple-peaked emission lines. The infrared colors show that the companion star is not a red giant, but may be a late main-sequence star or a subgiant. With the information available, V1721 Aql appears to be a good RN candidate.

\subsection{V1330 Cyg}
\label{sub:v1330cyg}
V1330 Cyg has been proposed as a candidate RN based on it light curve and amplitude \citep{rosenbush1999a}.  However, it has an S class light curve \citep{strope2010a}, the IR colors show no post-main-sequence star companion, and $A-A_0=3.6$, which, when combined, point to a likely CN.

\subsection{V2491 Cyg}
\label{sub:v2491cyg}
V2491 Cyg (Nova Cyg 2008b) displays a remarkable cusp in its eruption light curve; it is the first known example of a C class nova, a class for which even now only three members are known \citep{strope2010a}. \citet{tomov2008a} proposed V2491 Cyg as an RN based on vague similarities in its spectrum to those of two known RNe. Such arguments without any theoretical support are weak, especially since they could have alternatively pointed to CN examples. Indeed, \citet{naik2009a} point out various substantial differences between the spectra of V2491 Cyg and RNe. Among the indicators discussed in this paper, we have an inconclusive situation. The FWHM of the H$\alpha$ line is very large (4800 km s$^{-1}$; \citealp{naik2009a}) and the decline is fast ($t_3 = 16$ days; \citealt{strope2010a}), both of which suggest an RN. The other characteristics of V2491 Cyg, however, indicate that it is not an RN, as it has an orbital period of ${\sim} 0.1$ days \citep{darnley2011a}, an amplitude of 10 mag \citep{ribeiro2011a}, and a confusing situation among the emission lines, with a high iron line of Fe II \citep{lynch2008a} but possible (blended) He II \citep{helton2008a}. A model for this nova has produced an estimated WD mass of $1.3 \pm 0.02 M_\odot$ \citep{hachisu2009a}. The cusp light curve is unlike all known RN light curves and the implication is that substantially different physics is occurring \citep{hachisu2009a}, so it is unclear whether an RN is even possible. V2491 Cyg presents a conflicted case.

\subsection{KT Eri}
\label{sub:kteri}
KT Eri (Nova Eri 2009) has an outburst amplitude close to 7 mags and a fast decline ($t_3=10$ days; \citealp{hounsell2010a}). The FWHM of H$\alpha$ is 3400 km s$^{-1}$ \citep{yamaoka2009a} and H$\alpha$ shows a triple-peaked line structure (F. Walter 2010, private communication). The system may have a red giant companion star, based on its claimed 737 day photometric orbital period \citep{jurdana-sepic2012a}. KT Eri is a good RN candidate.

\subsection{HR Lyr}
\label{sub:hrlyr}
\citet{shears2007a} point to HR Lyr as being a possible RN based on its modest amplitude (9.5 mag), `fast' $t_3$ of 93 days, a hypothetical orbital period of 0.1 days, and an absolute magnitude in quiescence of +2.6 mag.  Unfortunately, the decline is slow by RN standards, and the star lies 3.7 mag above the dividing line on the amplitude/$t_3$ plot.  Additionally, even if there were a reasonable case for a 0.1 day orbital period, it would indicate a CN.  Finally, the absolute magnitude in quiescence is inside the range for CNe, and it is based on a dubious distance derived from the decline rate relations that are known to often be very wrong for RNe \citep{schaefer2010b}.  So we do not agree with the arguments of \citet{shears2007a}.  Nevertheless, J. Thorstensen (2009, private communication) reports a tentative photometric periodicity of 0.91 days, although he tells us that there is a substantial amount of uncertainty.  With this, there is a weak case for HR Lyr being an RN.

\subsection{BT Mon}
\label{sub:btmon}
\citet{surina2011a} selected a number of old novae based on their amplitude and $t_3$ and then obtained optical and IR colors for each system. They point to the old nova BT Mon as a possible RN of the U Sco type. BT Mon's relatively long (for a CN) orbital period (0.334 days, \citealp{downes2001a}) is also interesting. With our two new RN criteria (broad H$\alpha$ and high excitation lines), the 2010 km s$^{-1}$ FWHM of H$\alpha$ and presence of He II \citep{sanford1940a} are also indicative of a short recurrence time. Despite these suspicions, however, BT Mon has two properties that strongly indicate it is not recurrent: its 182 day $t_3$ and its F class (flat top at maximum) light curve \citep{strope2010a}. There are two additional measured physical quantities that strongly show that BT Mon is not recurrent: the mass of the WD is $1.04 \pm 0.06 M_\odot$ \citep{smith1998a}, which is too low for a short recurrence time, and the mass ejected by the 1939 eruption was $0.00003 M_\odot$ \citep{schaefer1983a}, which is too high to occur in a system with a short recurrence time. Additionally, a combination of archival data as well as magnitudes from the literature \citep{collazzi2009a,strope2010a} demonstrate that BT Mon had no undiscovered eruptions in the 122 years from 1890 to 2012. In all, we have strong reasons to reject BT Mon as an RN. This demonstrates that suspicions based on even three positive indicators can be incorrect.

\subsection{V2487 Oph}
\label{sub:v2487oph}
V2487 Oph (Nova Oph 1998) was proposed as an RN candidate by \citet{hachisu2002a} due to its very rapid decline and light curve plateau. In a detailed eruption model they derived an ejecta mass of ${\sim}6 \times 10^{-6} M_\odot$ and a WD mass of $1.35 \pm 0.01 M_\odot$, both of which are strong indicators of an RN. In addition, we noted in 2009 that V2487 Oph was in the RN region of the amplitude/$t_3$ diagram (Section \ref{sub:ampt3}), that it had a record high FWHM of 10,000 km s$^{-1}$, and prominent He II emission lines \citep{filippenko1998a,lynch2000a}. As such, V2487 Oph was an excellent RN candidate. On this basis, we made V2487 Oph the first target in our exhaustive search for previous eruptions in the Harvard and Sonneberg archival plate collections. We found a 1900 eruption in the Harvard collection, making V2487 Oph the tenth confirmed galactic RN \citep{pagnotta2009a}. This ideal case represents one of the main goals of the work in this paper, wherein we identify good candidates and then prove their RN nature by finding additional eruptions.

\subsection{V2672 Oph}
\label{sub:v2672oph}
V2672 Oph (Nova Oph 2009) is another nova with heavy extinction and relatively little observational data. \citet{munari2011a} note many RN properties and claim that V2672 Oph is a clone of U Sco. The nova is nearly the fastest known, with $t_3=4.2$ days, and has a near-record FHWM of 8000 km s$^{-1}$ for H$\alpha$ \citep{ayani2009a}; the H$\alpha$ line also shows the triple-peaked line structure, and He II lines are present \citep{schwarz2009a}. The light curve apparently plateaus at roughly 6 mags below peak. The lack of any 2MASS counterpart indicates that the secondary star is not a red giant. In all, four positive indicators (two of which are extreme) point to V2672 Oph being a strong RN candidate.

\subsection{GK Per}
\label{sub:gkper}
GK Per (Nova Per 1901) is notable as the first known and best example of a number of classic CN phenomena: periodic oscillations in the outburst light curve (O class), an expanding nova shell, a light echo, and subsequent dwarf nova eruptions. Nevertheless, there has long been a persistent suspicion (e.g. \citealp{szkody1994a}) that GK Per might be an RN because it has an evolved companion star (based on IR colors and $P_\mathrm{orb}$=1.9968 days; \citealp{szkody1994a}, \citealp{downes2001a}). This suspicion is increased by the Hubble Space Telescope observation of its clumpy nova shell \citep{shara2012a}, a property that is unique among novae except for the RN T Pyx \citep{schaefer2010a}. Nevertheless, GK Per is certainly not recurrent. In the 124 years from 1890 to the present it has been very well observed, and there has been only one nova eruption. Additionally, the mass of the WD has been measured to be $0.9 M_\odot$ \citep{crampton1986a}, $\geq 0.87 \pm 0.24 M_\odot$ \citep{morales-rueda2002a}, and $1.15 \pm 0.1 M_\odot$ \citep{hachisu2007a}; such low masses indicate that the recurrence timescale must be very long. This demonstrates that the presence of a post-main-sequence companion does not guarantee an RN.

\subsection{V445 Pup}
\label{sub:v445pup}
The unique Helium nova V445 Pup was observed to have strong IR dust emission roughly one month after its maximum, which suggested that the dust was pre-existing, from a previous eruption not too far in the past, indicating that V445 Pup is an RN \citep{lynch2001b}. However, the subsequent development of the light curve (with a deep and long-lasting minimum) demonstrated that the dust was formed in the current eruption, so the only evidence of the recurrent nature of V445 Pup is no longer valid.

\subsection{V1017 Sgr}
\label{sub:v1017sgr}
V1017 Sgr has had four eruptions observed, in 1901, 1919, 1973, and 1991. The 1919 eruption was longer and brighter than the other three events, with a light curve typical of a standard CN eruption. The other three small eruptions are typical dwarf nova eruptions, and the two most recent have been spectroscopically confirmed as such \citep{vidal1974a,webbink1987a}. This system is similar to GK Per, wherein a single CN event is surrounded by dwarf nova eruptions, however it is the only currently-known system to have a dwarf nova eruption observed prior to a regular nova outburst.

\subsection{V4444 Sgr}
\label{sub:v4444sgr}
V4444 Sgr has also been suggested to have a pre-existing circumstellar dust cloud, based on spectropolarimetry \citep{kawabata2000a}. The lack of a detectable dip in the light curve argues against the dust being created in the current eruption \citep{venturini2002a} and points to it being an RN \citep{kato2004a}. This is a plausible argument, although the lack of a model for the pre-existing dust cloud gives less confidence in its RN status.  In addition, a plateau in the light curve was reported, based on two isolated magnitudes \citep{kato2004a}, with this pointing to its RN status.  However, \citet{strope2010a} have greatly extended the light curve and all evidence for a plateau disappeared.  The light curve is very fast ($t_3 = 9$ days; \citealp{strope2010a}), a characteristic of most RNe.   Nevertheless, V4444 Sgr displays no high excitation lines in the outburst spectrum and the FWHM of the H$\alpha$ line is only 800 km s$^{-1}$ \citep{garnavich1999a}; such low energy properties are exhibited by none of the known RNe.  In all, we have a situation with conflicting indicators.

\subsection{NSV 1436}
\label{sub:nsv1436}
NSV 1436 (Ross 4) is generally fainter than $B=15$, but occasionally brightens to $B=13$ and brighter, leading \citet{brown2010a} to suggest that it is possibly an RN. Recent AAVSO monitoring, however, has revealed two eruptions from $V=16.5$ to $V=13$ within the past year, so NSV 1436 is certainly an ordinary dwarf nova.

\section{Recurrent Nova Candidate Search}
\label{sec:candidates}
We compiled the criteria described in Section \ref{sec:criteria} for both the RNe and the CNe; the compilations can be seen in Tables \ref{tab:rntable} and \ref{tab:bigtable}, respectively. Tables \ref{tab:rntable} and \ref{tab:bigtable} are nearly identical in structure. After the name of the system, the outburst characteristics are given: amplitude (in mags), $t_3$ (in days), $A-A_0$ (in mags), the FWHM of H$\alpha$ (in km s$^{-1}$), whether there is early He II, the highest reported Fe line, and the light curve class. Following these, the quiescence characteristics are listed: the orbital period (in days), the $J-H$ and $H-K$ colors, and whether it is in the red giant region on the $J-H$ vs. $H-K$ color-color diagram (Figure \ref{fig:ircolors}). The next column lists other indicators such as WD mass (Section \ref{sub:wdmass}), including those published by other groups. The only structural difference between Tables {\ref{tab:rntable}} and {\ref{tab:bigtable}} is the addition of an extra column in Table {\ref{tab:bigtable}}, denoted ``Category", which lists the ``likelihood of being an RN" category to which we have assigned that system, which is described in detail in Section {\ref{sub:fractionfromcn}}.

Using Table \ref{tab:bigtable}, we can easily identify the Galactic CNe which have multiple characteristics indicative of a short recurrence time; these are our RN candidates. The more RN-like characteristics a CN system shows, the more likely it is actually an unrecognized RN. We do not have complete information for every system, and even for those that we know a lot about, there are occasionally still uncertainties and/or contradictions. Additionally, the RNe and CNe do not fall into a neat bimodal distribution, but fall along a somewhat messy continuum. Because of these factors, there is no cut-and-dry rubric that can be used to classify the novae; instead, we have considered each system and the available characteristics individually.

Our good candidates are presented in Table \ref{tab:candidates}. This list includes seven systems from Table \ref{tab:bigtable} as well as three novae which erupted after the 2006 cut-off for Table \ref{tab:bigtable}. Our best candidate systems are V1721 Aql, DE Cir, CP Cru, KT Eri, V838 Her, V2672 Oph, V4160 Sgr, V4643 Sgr, V4739 Sgr, and V477 Sct.

\section{Recurrent Nova Fraction in Classical Nova Lists}
\label{sec:fraction}
Knowledge of the number of currently-listed CNe which are actually RNe, called the RN fraction $F_\mathrm{RN}$, is crucial for ascertaining the number of RNe in the Milky Way, for answering the question of whether the RN death rate equals the SN Ia rate, and thus for deciding whether RNe are acceptable SN Ia progenitor candidates. The following subsections describe the three independent methods we used to estimate the Galactic RN fraction. We stress that $F_\mathrm{RN}$ is the fraction of currently-known CNe which are actually RNe that have had multiple eruptions within the past century, not the fraction of the total CN population, the majority of which have never been identified as CNe. 

\subsection{RN Fraction From CN Properties}
\label{sub:fractionfromcn}
The first way to estimate the RN fraction is to closely examine the known CNe and identify the percentage that are likely RNe. For this analysis, it is vital to have an unbiased set of novae that were selected without any regard to their possible recurrent nature. We use the time-limited sample presented in Table \ref{tab:bigtable}, which consists solely of the CNe identified in \citet{downes2001a}, which was frozen on 2006 February 1. Some novae which erupted after that date are included in Table \ref{tab:candidates} and therefore should not be used for statistical analysis as they have been selected because they are good candidates.

With the collected data from Tables \ref{tab:rntable} and \ref{tab:bigtable} and our criteria from Section \ref{sec:criteria}, we can place each of our 247 novae into one of six categories, labeled A-F. Category A contains the known RNe, for which multiple eruptions have been confidently observed, listed in Table \ref{tab:rntable}. Category B contains strong RN candidates, for which many of our indicators strongly point to the system being recurrent, with only one eruption observed thus far, listed in Table \ref{tab:candidates}. Category C contains likely RN candidates, for which our evaluation of the characteristics in Table \ref{tab:bigtable} indicates that the probability is $\geq 50\%$ that the system is recurrent. Category D contains likely CNe, despite the system perhaps showing some small number of positive RN indicators. Category E contains systems which are certainly CNe based on the indicators in Table \ref{tab:bigtable}. In particular, systems showing any the following properties were placed into D or E: $\Delta A > 5.0$ mag; FWHM of H$\alpha < 1500$ km s$^{-1}$; neither He II nor high-excitation iron lines when both would have been visible; D, J, or F light curve classes; and measured $M_\mathrm{WD} < 1.1 M_\odot$. Category F contains systems for which there is not enough information to determine the status of the system. Table \ref{tab:rnfraction} presents the number of systems in each of the six categories, and the category for each nova is listed as the final column in Table {\ref{tab:bigtable}}. The fraction of RNe in published CN lists is then $F_\mathrm{RN}=(B+C)/(B+C+D+E)$. The uncertainty is $[F_\mathrm{RN} \times (1-F_\mathrm{RN})/(B+C+D+E)]^{0.5}$.

Classification was done independently by both authors of this manuscript for each nova system, taking into account all of the properties described in Section \ref{sec:criteria} as well as our knowledge of individual systems such as that described in Section \ref{sec:previous}. After we classified the full set of systems, we compared our results and found that they were almost completely in agreement. For the novae on which we initially disagreed, we examined them in more detail and agreed upon a classification after discussing the individual system characteristics. (As it is potentially sociologically interesting, we note that Pagnotta was more likely to classify a system as a possible RN, while Schaefer was generally more inclined to classify a system as a CN.) We were unable to come up with a strict, fully unambiguous decision tree to use for this classification, and as such there are likely unquantifiable uncertainties and possibly a small amount of bias included in our classifications, but we posit that they are reasonable, and attempt to describe our process here in as transparent a way as possible.

To start with, all known RNe were immediately placed into category A. All systems with zero usefully-observed properties were placed into category F, as were most systems with only one known characteristic. For some systems with only one characteristic, the value was strong enough that we could place them into category D, e.g. V720 Sco, which has no other data except spectra near peak, and those spectra show no He II and only Fe II, so it is very likely not recurrent. For one one-characteristic system, LZ Mus, the only known quantity is that it has a Plateau in its light curve, which is a decent RN indicator, so we classified it as C. For systems with two or more known characteristics, a balancing act starts to happen, which is not entirely objective, due to the nature of the systems and the characteristics. For example, HS Sge shows a plateau in its light curve, which for LZ Mus was enough by itself to kick the system up to C, but HS Sge also has $A-A_0 = 4.95$, greater even than the RN oddballs T Pyx and IM Nor, and therefore we bumped HS Sge back down to D, a likely CN, feeling that $A-A_0$ is somewhat stronger than light curve class because it is more physically understood and, in a sense, combines two characteristics. A counter-argument can of course be made, as it possibly can with many of the C and D category novae, especially for systems with many known characteristics which are conflicting. For example, V382 Vel has a marginally high expansion velocity (FWHM H$\alpha$ = 2390 km s$^{-1}$) and a high mass WD (M$_\mathrm{WD} = 1.23 \pm 0.05$M$_\odot$), but it has a high $A-A_0$ (4.31), no high excitation lines, an S-class light curve, and a very short orbital period (0.158 days). High expansion velocity and high mass WDs are very reliable indicators, but since the expansion velocity is only barely above the 2000 km s$^{-1}$ threshold and mass measurements are inherently uncertain, in this case we weight the other characteristics more and classify V382 Vel as a category D; in general, if fewer than half of the known indicators point to an RN, we probably classify that system as D or E. However, for V4633 Sgr, which also has only two of six characteristics that fall in the RN regime, since those two characteristics are both very strong (early He II high excitation lines and a P-class light curve), and some of the others are relatively close to the RN thresholds (e.g. FWHM H$\alpha$ = 1600 km s$^{-1}$), we categorize V4633 Sgr as a C class, or likely RN. Of course, for systems where all or almost all of multiple characteristics indicate a short recurrence time, it is easy to categorize that system into B, a probable RN, such as DE Cir, which has four of its five characteristics point to it being a hidden RN. This holds true except for systems about which we have outside prior knowledge that supersedes the characteristics in {\ref{tab:bigtable}}. For example, GK Per looks promising due to its long orbital period and likely high mass WD, but it has  been very well observed since its 1901 nova eruption and has not been seen to erupt again, instead showing a number of dwarf nova outbursts, which are caused by a different physical mechanism (instabilities in the accretion disk) and indicate a large change in the mass accretion rate, therefore GK Per is in category E. This process is highly system-dependent and likely has unquantified uncertainties, but again we are confident that it is robust and useful for the following analysis.

Category A systems are listed in Table \ref{tab:rntable}, and Category B systems are listed in Table \ref{tab:candidates}. The Category C systems are V368 Aql, V606 Aql, V1229 Aql, V1493 Aql, DD Cir, V693 CrA, V2275 Cyg, Q Cyg, DN Gem, HR Lyr, LZ Mus, CP Pup, V574 Pup, V1172 Sgr, V1275 Sgr, V4327 Sgr, V4361 Sgr, V4633 Sgr, V4742 Sgr, V4743 Sgr, V696 Sco, V697 Sco, V723 Sco, V1141 Sco, V1187 Sco, V1188 Sco, X Ser, CT Ser, and QU Vul.

A concern is that our sample of 247 systems may have a bias because many of the novae have relatively little data\textemdash even beyond our exclusion into Category F\textemdash and a poorly-observed nova is less likely to be recognized as an RN. To address this, we examine the subset of well-observed novae in \citet{strope2010a}. This sample has thorough light curve coverage, and generally good spectroscopic and quiescence observations as well. The statistics for the Strope sample are also presented in Table \ref{tab:rnfraction}.

The two samples give RN fractions of 24.3\%$\pm$3.5\% and 24.1\%$\pm$4.8\%, so we take $24\% \pm 4\%$ as the final result for this first method. Since there are approximately 400 known CNe to date, ${\sim} 96$ of them should be RNe. The ratio of unidentified RNe to the known RNe is therefore ${\sim} 10:1$.

\subsection{RN Fraction from Number of Known RNe}
\label{sub:fractionfromrne}
There must be some proportionality between the number of known RNe and the number of RNe masquerading as CNe. The fact that there are ten known RNe can thus be used to estimate the number of hidden RNe and therefore the RN fraction. To do this, we must have a good understanding of the discovery efficiencies for individual nova events. If the probability of discovering an individual nova eruption is low, there must be a high RN fraction to produce the ten known RNe. If the discovery efficiency is high, the RN fraction should be relatively low. To recover the RN fraction, we need a detailed model that examines the discovery efficiencies for novae as a function of the individual system properties.

We have previously studied the discovery efficiencies of nova events extensively 
\citep{pagnotta2009a,schaefer2010b}. This was done in part using our detailed listings of plate times and depths for many novae, for many years, for both the Harvard and Sonneberg plate archives. We also used our comprehensive listings of search epochs, cadences, and magnitude limits for both amateur and professional nova searches conducted by many people and groups throughout the world. The results are detailed and include exact knowledge of the sizes of the solar, lunar, and observational gaps, i.e. the fraction of time over which someone could and would have discovered a nova eruption. We have pieced this together to derive a relatively simple formula that quantifies the discovery efficiency for an undirected nova search picking up an eruption \citep{schaefer2010b}. This discovery efficiency is 
\begin{equation}
F_\mathrm{disc}=f_\mathrm{disc}(V_\mathrm{peak}) \times 0.67 \times (t_3/44).
\end{equation}
The factor $f_\mathrm{disc}(V_\mathrm{peak})$ is the discovery efficiency for a $t_3=44$ day nova that is well placed in the sky, and its values are 1.0 for $V_\mathrm{peak}$=2 mag or brighter, 0.35 for $V_\mathrm{peak}$=4 mag, 0.22 for $V_\mathrm{peak}$=6 mag, 0.14 for $V_\mathrm{peak}$=8 mag, and 0.09 for $V_\mathrm{peak}$=10 mag. We take a linear interpolation between these values and assume the efficiency falls linearly to zero at $V_\mathrm{peak}$=16 mag. We note that we do treat the $V_\mathrm{peak}$ and $t_3$ as independent variables despite their possible relationship via the Maximum Magnitude Rate of Decline (MMRD) relationship \cite{della-valle1995a}; recent work by \citet{kasliwal2011a} has raised questions about whether or not the MMRD is as tight a correlation as previously thought, so we choose to take the conservative route and keep them independent for the purposes of this calculation.

The discovery of the first eruption of a system is always from an undirected search. The discovery of the second eruption is generally also from an undirected search. Once a system is known to be an RN, then {\it directed} searches are made, both ongoing into the future as well as with archival data into the past. The directed discovery efficiency is higher than the undirected discovery efficiency, and can in fact be quite high. For example, U Sco had a high directed discovery efficiency in the several years leading up to its 2010 eruption because the prediction of its eruption led to intense surveillance by many professional and amateur astronomers, with the amateurs able to push the monitoring deep into the solar gap. Here, it only matters that we have a fairly accurate measure of the undirected discovery efficiency for a known $V_\mathrm{peak}$ and $t_3$.

For each of the novae in Table \ref{tab:bigtable}, we have a known peak magnitude, and 72\% have a known $t_3$.  Those that do not have a $t_3$ are assigned the value of 44 days, which is the median value for CNe.  In addition, we must handle the 147 novae that are not listed in Table \ref{tab:bigtable}, including 91 poorly observed eruptions that are in \citet{downes2001a} but not Table \ref{tab:bigtable} as well as 56 systems which erupted after the stop date of the \citet{downes2001a} catalog. For these, we have cloned the parameters for the first 147 novae in Table \ref{tab:bigtable}.  For the 10 known RNe, we use their parameters from Table \ref{tab:rntable}.  With this, we can calculate the probability that any one nova event will be discovered for each of the 394 novae currently known.

We can now simulate with a Monte Carlo analysis the discoveries of all the novae and determine which systems have one eruption discovered and which systems have more than one discovered.  This simulation has only one free parameter: the fraction of the discovered systems that are actually RNe with multiple eruptions.  There are large populations of recurrent and non-recurrent systems which have had at least one eruption since 1890.  Out of the many thousands of eruptions, only  a small fraction are discovered, and this leads to the 394 systems currently known.  Of these 394 systems, some fraction are recurrent and have had multiple eruptions since 1890.  Of these RNe, most have had only one eruption discovered, and a few have had two or more eruptions detected.  Our simulation models all these numbers based on the RN fraction input.  If the RN fraction is near zero, then we would recognize near zero RNe, while if the RN fraction is large then we would recognize many more than ten RNe.  For a given RN fraction, we run the simulation many times and calculate the number of discovered RNe, then average these numbers and calculate their RMS scatter.  We then vary the RN fraction until the number of discovered RNe is 10, exactly as is observed out of the 394 systems known to date.  The 1$\sigma$ uncertainty on this derived RN fraction is the range over which the observed number (10) is within one RMS of the average number for that assumed RN fraction.

In each Monte Carlo simulation, each star is assigned a random number that determines whether it is recurrent based on the RN fraction.  The RNe are then assigned a randomly selected recurrence time scale ($\tau_\mathrm{rec}$) from a distribution.  The adopted distribution is flat in $\log [\tau_\mathrm{rec}]$ from 10 to 100 years, which is a reasonable match for the ten known RNe, after correction for selection effects.  For the simulated $\tau_\mathrm{rec}$ and a random phase in this recurrence cycle in the year 1890, the number of eruptions from 1890 to 2012 is calculated.  For each of these eruptions, a random number is compared to the undirected search efficiency to decide whether that particular eruption is `discovered'.  The total number of discovered eruptions is then tallied up.  If this is zero, the simulation is repeated until a non-zero number of eruptions for that nova is found.  We then get the total number of detected eruptions from all of the CNe (one each) and all of the simulated RNe (one or more each).  From this list, we then count the number of discovered RNe (i.e., those recurrent systems that have two or more simulated eruptions discovered).  Such simulations are repeated many times for the same set of inputs to beat down the usual statistical variations.  These variations are important because they also show the typical variations that we should expect for the nova population that is actually realized in the sky.

The result of this Monte Carlo analysis is the average number (and its RMS scatter) of detected RNe as a function of the RN fraction.  With this, we find that the population of 394 known nova systems will produce 10 known RNe for an RN fraction of 12\%$\pm$3\%. 

\subsection{RN Fraction From RN Discovery Efficiency}
\label{sub:fractionfromdisceff}
The RN fraction can also be determined by the discovery efficiency of each of the ten known RNe. If it is highly probable that an individual RN will be recognized as recurrent, then most of them will have already been discovered and few will be lurking among the CN lists. If the identification of two or more eruptions occurs with low probability, then there must be many RNe for which only one eruption has been detected. Our third method takes the discovery efficiency for each of the known RNe and deduces how many similar systems must exist to allow for the one system recognized as an RN. We can also calculate how many of these similar systems will have only one discovered eruption, and hence how many RNe are masquerading as CNe.

Full discovery details, undirected discovery efficiencies ($F_\mathrm{disc}$), and recurrence time scales ($\tau_\mathrm{rec}$) are known for all ten RNe \citep{schaefer2010b}. Values for $F_\mathrm{disc}$ and $\tau_\mathrm{rec}$ are taken from Tables 18 and 21 of \citet{schaefer2010b} and summarized in our Table \ref{tab:rndiscovery}. Given that the first discovered eruption is always from an undirected search and that few CNe are monitored with any useful directed nova search even for short times, the relevant efficiency of the discovery of RNe is that of undirected searches in general.

The discovery statistics for systems like the known RNe can be calculated using another Monte Carlo simulation. For each simulation, for each RN, the number of eruptions from 1890 to 2012 is determined by the given $\tau_\mathrm{rec}$ and a randomly chosen phase within this recurrence cycle. For each eruption that occurs, a random number is selected for comparison with the undirected discovery efficiency to ascertain whether the eruption is identified. The number of discovered eruptions is tallied for each RN and simulation. We tabulate the fraction of the time that zero eruptions are discovered ($f_0$), one eruption is discovered ($f_1$), and two or more eruptions are discovered ($f_{\geq 2}$) for 10,000 simulations and report these in Table \ref{tab:rndiscovery}.

Since most RNe are never identified, there must be many in existence for any one of the known RNe to be recognized. We can consider the example of RS Oph as an illustration. For 100 RS Oph-like systems, 58 would never be seen to erupt, 33 would have one discovered eruption and be classified as CNe, and 9 would be recognized as RNe since multiple eruptions would be discovered. To turn it the other direction, since we observe one RS Oph system, there should be 11 such RNe currently active, with ${\sim} 6$ of those never seen, ${\sim} 4$ identified as CNe, and the 1 recognized RN. In general, for every known RN, there should be $N_\mathrm{CN}=f_1/f_{\geq 2}$ apparent CNe recorded. The required number of false-CNe is tabulated for each RN in Table \ref{tab:rndiscovery}, rounded to the nearest integer. Additionally, for every known RN, there must be roughly $N_\mathrm{RN}=1/f_{\geq 2}$ other RNe in existence.

We must be careful to which known RNe we apply this logic. The original calculation assumes that {\it both} of the first two eruptions were discovered with undirected nova searches during the years 1890-2012. This assumption is incorrect for several of the known RNe. V2487 Oph's first eruption (in 1998) was discovered by an amateur during an undirected search \citep{nakano1998a}, but its RN-like properties led us to perform a directed search to find the second-discovered 1900 eruption. Similarly, the second-discovered eruptions of T Pyx and U Sco were found during directed searches \citep{leavitt1913a,thomas1940a}. For T CrB, L. Peltier followed up the 1866 eruption with a long-running directed search, but this does not affect the statistics because the 1946 eruption was discovered by many independent amateurs during undirected searches. For T CrB, the two discovered eruptions are not from 1890-2012, since the first one appeared in 1866. In all four of these cases, there need be substantially fewer similar hidden RNe to produce the ones that we know.

For the three known RNe for which the second discovery came during a directed search, the probability of detecting the second eruption increases significantly since directed searches are much more efficient. \citet{schaefer2010b}'s Table 20 quantifies the directed search efficiencies for each known RN for each year after 1890. For T Pyx, the probability of the second discovery increases from 0.19 to near unity. For V2487 Oph, the probability goes from 0.012 to 0.33. For U Sco, the probability increases from 0.006 to 0.43. For each of these cases, the fractions can be calculated analytically. For T CrB, we take the probability of discovering an eruption to be 10\% from 1850 to 1890, based on broad experience with historical variable star work, because detailed lists of observing epochs and depths are unavailable.

With this, the sum over $N_{CN}$ for all ten RNe is 139; these 139 so-called CNe are actually RNe for which only one eruption has been discovered. The sum over $N_{RN}$ is 1604; there are currently roughly 1600 active RNe in our galaxy that are bright enough to be detected.  Due to the low discovery efficiencies, this large number is required to produce the ten known RNe, and there are likely many more spread throughout the Milky Way far outside the local neighborhood of visibility.

The Monte Carlo calculation can also be used to estimate the uncertainties.  The 10,000 Monte Carlo runs allow for a pretty good estimate of $N_{RN}$ and the probabilities of discovering eruptions (as given in Table {\ref{tab:rndiscovery}}), with small errors introduced by the Monte Carlo random statistics.  But in any one run from 1890-2012 (including the one realization that is the real historical case), the numbers of single and multiple discoveries will vary randomly around these averages.  For the calculated number of systems like each known RNe and the discovery probabilities, we can calculate the RMS scatter in the number of single and multiple detected eruptions (for binomial statistics) and propagate these uncertainties through to the number of RNe masquerading as CNe.  For the example of CI Aql, the Monte Carlo gives $f_{\geq 2}=4\%$, so the number of similar RNe in the sky (with $F_{disc}\sim0.05$ and $\tau_{rec}\sim24$ years) must have $N_{RN}\sim28$ with 23\% having single detected eruptions for an average of $N_{CN}=7$ (see Table {\ref{tab:rndiscovery}}), but this last number should randomly vary with an RMS of $[28 \times 0.23 \times (1-0.23)]^{0.5}$ or $\pm$2.2.  Each RN will have an uncertainty in $N_{CN}$ associated with the randomness of discovery, and these can be propagated to the total number of RNe masquerading as CNe.  In all, the uncertainty on our derived 139 masquerading RNe is $\pm11$.

This calculation of $N_{RN}$ is essentially asking how many systems just like each known RN (i.e., with the same peak magnitude, $t_3$, and $\tau_{rec}$) are out there such that just one of them will be discovered.  With this, we are taking the ten known galactic RNe to span the parameter space for RNe, and with appropriate correction for discovery probabilities, to plausibly represent the distributions of peak magnitude, $t_3$, and $\tau_{rec}$.  With only ten galactic RNe, we can have only poor resolution of the underlying distributions.  Some RNe (like T Pyx) have a high probability of being detected as a recurrent system, so there can be few similar systems up in the sky other than the one discovered (T Pyx itself).  Other RNe (like V394 CrA) have such a low discovery probability that there must be many similar systems up in the sky for one to be discovered.  If two of some type are discovered (like V745 Sco and V3890 Sgr), then the calculation will have the correct numbers because both known systems will be used.  A large possible problem comes in if there are some types of RNe with low $F_{disc}$ that have zero discovered systems, because then entire classes will not be represented in our calculation.  To take two examples, there is likely a large population of RNe at the far side of our Milky Way that are so faint that they will not be discovered, and there might be a population with $t_3\sim3$ days and $\tau_{rec}\sim80$ years for which none have been discovered.  The first example makes us realize that our Monte Carlo calculation is {\it not} giving us the total RN number in our galaxy, but rather it is merely correcting for discovery inefficiencies such that some valid extrapolation to the whole Milky Way is still needed.  The second example makes us realize that we could well be missing large numbers of RNe that are hard to detect, and so our Monte Carlo calculation really just provides a lower limit.  However, for the question of the number of RNe masquerading as CNe, these hard-to-discover classes will likely contribute only a negligible number of systems with only one eruption, for the very reason that they are hard to discover.  In all, our Monte Carlo calculation of the $N_{CN}$, makes a plausible coverage of the distributions of RN properties, even if the number might be larger by some unknown amount.
 
We do not have any high confidence in the formal uncertainty in $N_{CN}$ of $\pm$11.  Part of the reason is that the various inputs to our calculation have substantial and unknown uncertainties.  For example, the recurrence time scale of CI Aql could be twice as large with an unknown likelihood, and the $F_{disc}$ values change over the decades in ways difficult to quantify.  Part of the reason is that the ten known RNe provide a poor-resolution sample of the real underlying distribution of $F_{disc}$ and $\tau_{rec}$, with any correction to the unknown real distributions being unknown.  Part of the reason is that there might be large systematic errors, for example caused by possibly large populations of RNe for which the types happened to have zero discovered to date.  With all of these problems, we guess that a better representation might be like $\pm$100.  But we cannot put forth any such guess with any justification or confidence, so we can only quote the formal $\pm$11 and note that this is likely a substantial underestimate.

Roughly 139$\pm$11 of the currently cataloged CNe are actually RNe.  But this is not out of the 247 novae listed in Tables 2 and 3, because these lists include neither the many faint novae with sketchy data, nor the novae discovered since the beginning of 2006. Indeed, the second eruptions of half of the known RNe (IM Nor, CI Aql, V394 CrA, V745 Sco, and V3890 Sgr) came as a complete surprise from 1987 to 2002, since the first eruption of each system had fewer than a dozen magnitudes and no spectroscopy. The poorly observed CNe provide a large pool of systems which could be RNe.  If we count all poorly-observed systems, as well as those featured in our Table 3, there are 338 novae in the \mbox{\citet{downes2001a}} catalog, as well as 56 novae which have been discovered since the beginning of 2006. Thus, the real pool of so-called CNe contains 394 systems, and the RN fraction is 139$\pm$11 out of 394 systems or 35\%$\pm$3\%, with this formal error bar likely being much too small.

\subsection{Final RN Fraction}
\label{sub:finalfraction}
We have calculated the Galactic RN fraction using three greatly different techniques. It is a challenge to come up with reliable estimates of the total uncertainty for these three measures of the RN fraction.  We report the statistical error bars, but they do not include any of a variety of systematic errors and biases that we can think of.  In the first method, we cannot translate the mixed sets of available data and the various fuzzy criteria into a probability that a given CN is really a RN, and as such cannot  derive the RN fraction with good accuracy.  We have already pointed to the possible bias caused by the poorly-observed events might have a higher, or lower, RN fraction, although our use of two samples (the 247 system time-limited sample and the Strope sample) suggests that this bias is negligible.  In the second method, the discovery efficiency is undoubtedly more complex than our formula allows, and some of our assumptions in the Monte Carlo calculation (for example, that second eruptions will only be discovered by undirected searches) might be imperfect.  And there might be biases, for example, with our setting $t_3$=44 days when the duration is unknown, which might be incorrect because RNe are faster than CNe.  In our third method, the underlying distributions of $F_{disc}$ and $\tau_{rec}$ might be substantially different than those derived from the known RNe, leading to errors in the RN fraction that cannot be calculated.  We also have the potentially substantial bias caused by possibly-common types of RNe (say, those with a very fast $t_3$ and a relatively long inter-eruption interval) having no observed examples, so these will not be included in the derived number of currently active RNe.  The readers can easily come up with other potential problems, some of which might be significant.  All this is to say that our three measured RN fractions have poorly known systematic uncertainties and might be subject to substantial unrecognized biases.

We have three estimates of the RN fraction as $24\% \pm 4\%$, $12\% \pm 3\%$, and $35\% \pm 3\%$.  The estimates appears to be significantly different from each other, but this is not the full story, because the quoted error bars are for the statistical errors alone\textemdash for the third value especially the quoted error is likely too small\textemdash and the real total error bars (including systematics) could easily make all three values consistent.  All three estimates are independent measures of the same quantity (the RN fraction) by greatly different methods with unknown and possibly substantial uncertainties.  In this situation, we can only use the three estimates to pick out a range that is reasonable, with the size of the range providing some sort of estimate of a final overall error bar.  This is not perfect, but it is the best that can be done, and it does have the advantage of giving an approximate answer provided that we do not insist on claiming high accuracy.  A straight average gives an RN fraction of 24\%.  It is reasonable to state that the RN fraction is roughly 25\%, and is likely somewhere between 15\% and 35\%.

We are not claiming any high accuracy to our result that $F_{RN}=25\% \pm 10\%$, because we know there are almost certainly large unknown systematic errors in our three estimates.  But our analysis from Section {\ref{sub:fractionfromcn}} does show that $F_{RN}\ll50\%$, because the majority of catalogued CNe are not RNe as shown by their white dwarf masses, their ejection velocities, and so on.  And critically, our analysis in Section {\ref{sub:fractionfromdisceff}} confidently shows that $F_{RN}\gg3\%$ (i.e., there are many more than the 10 known RNe hiding in the nova catalogues), because the number of hiding RNe is roughly $10/ \langle f_{\geq 2} \rangle$, and we know confidently that $\langle f_{\geq 2} \rangle \ll 1$.  Both of these extreme limits are robust, simple, and cannot be gotten around by unknown biases or systematic errors.  Given these two confident limits ($3\% \ll F_{RN} \ll 50\%$), our value of $F_{RN}=25\% \pm 10\%$ seems quite reasonable.
        
Again we stress that this RN fraction of approximately one-quarter is the percentage of cataloged CNe that are actually RNe. We can use the previously-discussed discovery statistics of novae to estimate the total number of CNe in our galaxy. There are ${\sim}400$ novae in our current catalogs. The average discovery efficiency for novae is 10\%, so we multiply $400 \times 10$ to get 4000 systems which have erupted in the past century. Most of the CNe will not have erupted during this time period, though. To account for this, we consider the average recurrence timescale of true CNe, which is estimated to be $10^4{-}10^6$ years \citep{truran1986a,ritter1991a,yaron2005a}. Then, during the past 100 years, only 0.01-1\% of CNe have had eruptions at all, so we multiply 4000 by $10^4/100$ to $10^6/100$ to get a total galactic CN population of somewhere between $4\times 10^5$ and $4\times 10^7$ systems.  Our $\sim$1600 RNe are thus only a fraction of a percent of the total number of CNe.

\section{Searches for Prior Eruptions}
\label{sec:searches}

Presented in this paper is a list of good recurrent nova {\it candidates}; we stress that a system cannot be definitively classified as recurrent until two or more eruptions are observed, traditionally within 100 years or so. There are two ways to find these eruptions: looking forward and looking backward. Looking forward requires frequent monitoring of candidates. It is currently not feasible to do this comprehensively for a large number of systems, but in the future, all-sky surveys with real time results (such as ASAS, PanSTARRS, and LSST) may turn this into a reasonable option. 

Looking backward takes advantage of the extensive astronomical plate archives at the Harvard College Observatory (HCO) in Cambridge, MA, and the Sonneberg Observatory in Sonneberg, Germany, which provide regular all-sky coverage dating back to 1890. HCO and Sonneberg have approximately 500,000 and 300,000 plates, respectively, and the collections complement each other nicely: HCO has better coverage of the early years (1890 - 1953) while Sonneberg has better coverage of the later years (1945 - present). At Sonneberg, they are still taking large-scale, all-sky survey data today, on films instead of glass plates. Instead of waiting many decades for the hopeful results of forward-looking surveys, we can look backwards in the archival plates for a 124 year history of eruptions.

Twenty RN eruptions have been observed on the plates at HCO and Sonneberg, including the discovery of the 1900 eruption of V2487 Oph, which was found during the course of this project and provided proof that V2487 Oph was in fact recurrent \citep{pagnotta2009a}. One strong candidate (V838 Her) has been essentially fully checked at both archives, and another (V2672 Oph) has been nearly fully checked at HCO, but no prior eruptions have been found. This is not unexpected, for the same reasons that contribute to the low overall discovery efficiencies for RNe in general, namely short eruption durations that can easily be missed during solar and lunar gaps, and times of less frequent sky monitoring. Four other good candidates are at various stages of being checked; although the procedure is simple, the process is time-consuming. A summary of our searches to date can be seen in Table \ref{tab:plates}, which gives a breakdown of which plate series have been checked, approximately how many plates were present for each series, the center location of the plate series (for the patrol plates), and a rough idea (in the footnotes) of what time period each series covers, for seven systems. The digitization projects that are currently in progress at HCO and Sonneberg will make it much easier to check all candidates for previous eruptions. We note that many Baade's Window plates had been pulled for scanning at HCO during recent trips there. We made every attempt to examine all of the pulled plates for the series listed as checked, however it is possible that some number were missed. 

\section{Implications}
\label{sec:implications}

We conclude that the Galactic RN fraction, the percentage of currently-labeled CNe that are actually RNe, is $F_{RN}=25\% \pm 10\%$. With this, we expect that roughly 100 (or between 60 and 140) of the 394 systems labeled as CNe are in fact currently active RNe for which only one eruption has been thus far discovered.  We can expect many more second eruptions to be discovered in the upcoming decades, with most of these coming from old, largely-ignored novae with scanty data.    

One implication is the imperative to seek second eruptions in archival data for the Category B systems listed in Table \ref{tab:candidates}.  When seeking new RNe, the fastest, most time-effective method is to search through old astronomical photographs now residing in the archives, most of which are at Harvard and Sonneberg.

Another implication is for nova researchers to realize that roughly a quarter of the systems they are studying are in fact RNe.  Hopefully, with our criteria, the likely recurrent systems can be picked out and recognized.  With this realization, models will be constructed differently, and anomalies might get explanations.  Additionally, some key questions might receive deserved attention, for example whether the RN candidates have WDs that are carbon-oxygen or oxygen-neon-magnesium types.

The most important implication is that there must be $\sim$1600 completely undiscovered RNe in the galaxy, which has vital implications for RN demographics.  The previous study \citep{della-valle1996a} did not account for discovery efficiencies, and therefore their RN death rate must be revised upward. Following the calculations and notation of \citet{della-valle1996a}, our new $N\approx 1600$, so the birth rate of SNe Ia from RNe (assuming they require ${\sim}(1{-}2) \times 10^6$ years to become SNe Ia) is then $(0.8{-}1.6)\times 10^{-3}$, which is one-quarter to one-half of the galactic SN Ia rate. Based on these rates, RNe are therefore viable candidates to contribute a significant fraction of the SNe Ia.

We note that demographics alone are not enough to prove that RNe are progenitors. There are at least two other physical conditions that must be met: the WDs in the systems must undergo a net mass gain across eruptions (i.e. they must accrete more during the inter-eruption interval than is blown off during the eruption itself) and they must be composed primarily of carbon and oxygen. There is preliminary evidence that at least some of the RNe do undergo a net mass gain (CI Aql; \citealp{schaefer2011a}), but others may have oxygen-neon WDs (U Sco; \citealp{mason2011a}), so the situation is unclear. Additionally, results of progenitor searches in nearby Type Ia supernova remnants exclude the possibility of RN progenitors most cases \citep{edwards2012a,gonzalez-hernandez2009a,li2011a,nugent2011a,ruiz-lapuente2004a,schaefer2012a}. Nevertheless, recent work indicates the likely presence of two distinct progenitor channels \citep{brandt2010a,greggio2010a}, so it is worthwhile to investigate all systems which may contribute significantly to the SN Ia rate. Our results here show that, when discovery efficiencies are properly considered, the RN population is large enough to provide up to one half of the SN Ia rate, and therefore it is important to continue studying these systems.

\acknowledgments
This research was supported by the National Science Foundation, the LSU Board of Regents, the LSU Graduate School, and the Kathryn W. Davis Postdoctoral Scholar program. We thank Alison Doane, Jaime Pepper, and the entire DASCH team at HCO (led by PI Josh Grindlay), and Peter Kroll, Klaus L${\rm \ddot o}$chel, and the rest of the 4$\pi$ Systeme staff at Sonneberg for allowing access to the plate stacks and for many helpful discussions. We also thank Limin Xiao for her assistance checking some of the plates for V838 Her listed in Table \ref{tab:plates}. This research has made use of the SIMBAD database, operated at CDS, Strasbourg, France. We acknowledge with many thanks the variable star observations from the AAVSO International Database contributed by observers worldwide and used in this research.

\newpage
\begin{deluxetable}{l c c}
\tablecaption{Efficiency of RN Identification Characteristics$^a$}
\tabletypesize{\scriptsize}
\tablewidth{0pt}
\tablehead{
	\colhead{Indicator}			&
	\colhead{\% of RNe$^b$}			&
	\colhead{\% of CNe}			}	
\startdata	
$A-[14.5-4.5\log(t_3)]<0$	&	78	&	2	\\
P$_\mathrm{orb} > 0.6$ days	&	78	&	13	\\
$J-H>0.7$ mag \& $H-K>0.1$ mag	&	44	&	15	\\
FWHM of H$\alpha > 2000$ km s$^{-1}$	&	100	&	53	\\
High-Excitation Lines (He II and Fe X)	&	100	&	21	\\
P class (plateau in light curve)	&	56-89	&	17	\\
$M_\mathrm{WD}>1.2 M_{\odot}$	&	100$^c$	&	26	\\
\enddata
\tablenotetext
{a}{Note that to avoid any biases, we excluded the recurrent nova V2487 Oph from all of the calculations that went into this table, as its recurrent nature was discovered by using a subset of the characteristics to identify it as a good candidate, after which we searched for and found a previous eruption in the historical record \mbox{\citep{pagnotta2009a}}.}
\tablenotetext
{b}{For clarity, we show the values here rounded to the nearest percent. Please see the corresponding text in Section {\ref{sec:criteria}} for higher precision numbers and errors on the percentages.}
\tablenotetext
{c}{This 100\% value for RNe is not a fully observed quantity, however theoretical models strongly require that all RNe have $M_\mathrm{WD}>1.2 M_{\odot}$ (e.g. \mbox{\citealp{yaron2005a}}), so we are confident that it is robust, despite being model-dependent.}
\label{tab:indicatorfractions}
\end{deluxetable}

\newpage
\begin{deluxetable}{l c c c c c c c c c c c c}
\rotate
\tablecaption{Properties of the Known Galactic Recurrent Novae}
\tabletypesize{\tiny}
\tablewidth{0pt}
\setlength{\tabcolsep}{0.04in} 
\tablehead{
	\colhead{Nova}						&
	\colhead{Amplitude (mag)}			&
	\colhead{t$_3$ (days)}				&
	\colhead{$A-A_0$}					&
	\colhead{FWHM H$\alpha$}			&
	\colhead{He II}						&
	\colhead{Highest Iron}				&
	\colhead{Light Curve}				&
	\colhead{P$_\mathrm{orb}$ (days)}		&
	\colhead{$J - H$}					&		
	\colhead{$H - K$}					&	
	\colhead{Red Giant?}				&	
	\colhead{Other}					}	
\startdata	
CI	Aql	&	7.7	[1]$^a$	&	32	[1]	&	-0.03	&	2250	[3]	&	Yes	[3]	&	Fe VII	[4]	&	P	[5]	&	0.618	[6]	&	0.51	[7]	&	0.22	[7]	&	N	[2]	&	...		\\
V394	CrA	&	11.2	[1]	&	5.2	[1]	&	-0.08	&	4600	[8]	&	Yes	[8,9]	&	...		&	...		&	1.516	[10]	&	...		&	...		&	N	[2]	&	...		\\
T	CrB	&	7.3	[1]	&	6	[1]	&	-3.70	&	4980	[11]	&	Yes	[11]	&	Fe X, XIV	[11]	&	S	[5]	&	227.569	[6]	&	0.94	[12]	&	0.26	[12]	&	Y	[2]	&	...		\\
IM	Nor	&	9.8	[1]	&	80	[1]	&	3.86	&	2380	[13]	&	Yes	[13]	&	Fe X	[13]	&	P	[5]	&	0.103	[6]	&	...		&	...		&	N	[2]	&	...		\\
RS	Oph	&	6.2	[1]	&	14	[1]	&	-3.14	&	3930	[14]	&	Yes	[15]	&	Fe X, XI, XIV	[16]	&	P	[5]	&	455.720	[6]	&	0.75	[12]	&	0.33	[12]	&	Y	[2,16]	&	...		\\
V2487	Oph	&	8.2	[1]	&	8	[1]	&	-2.64	&	10000	[18,19]	&	Yes	[18]	&	...		&	P	[5]	&	...		&	...		&	...		&	N	[2]	&	$M_\mathrm{WD}=1.35 \pm 0.01 M_\odot$	[20]	\\
T	Pyx	&	9.1	[1]	&	62	[1]	&	2.67	&	5350	[21]	&	Yes	[21]	&	Fe II	[21]	&	P	[5]	&	0.076	[6]	&	-0.02	[12]	&	0.45	[12]	&	N	[2]	&	...		\\
V3890	Sgr	&	7.4	[1]	&	14	[1]	&	-1.94	&	2140	[27]	&	Yes	[9]	&	Fe X	[9]	&	S	[5]	&	519.7	[10]	&	1.06	[12]	&	0.41	[12]	&	Y	[2]	&	...		\\
\enddata
\tablenotetext{a}{Reference Key:
	1 = \citet{schaefer2010b};
	2 = This Paper;
	3 = \citet{kiss2001a};
	4 = \citet{schmeja2000a};
	5 = \citet{strope2010a};
	6 = \citet{downes2001a};
	7 = \citet{harrison1992a};
	8 = \citet{sekiguchi1989b};
	9 = \citet{williams1991a};
	10 = \citet{schaefer2009a};
	11 = \citet{bloch1946a};
	12 = \citet{harrison1993a};
	13 = \citet{orio2005a};
	14 = \citet{bode2007a};
	15 = \citet{joy1961a};
	16 = \citet{dufay1964b};
	17 = \citet{szkody1994a};
	18 = \citet{lynch2000a};
	19 = \citet{filippenko1998a};
	20 = \citet{hachisu2002a};
	21 = \citet{catchpole1969a};
	22 = \citet{sekiguchi1988a};
	23 = \citet{anupama2000a};
	24 = \citet{hoard2002a};
	25 = \citet{sekiguchi1990a};
	26 = \citet{duerbeck1989a};
	27 = \citet{wagner1990a};
	28 = F. Walter, private communication
	}
\label{tab:rntable}
\end{deluxetable}

\newpage
\begin{deluxetable}{l c c c c c c c c c c c c c c c c c c c c c c c c}
\rotate
\tablecaption{Properties of the Known Galactic Classical Novae}
\tabletypesize{\tiny}
\tablewidth{0pt}
\setlength{\tabcolsep}{0.02in} 
\tablehead{
	\colhead{Nova}						&
	\colhead{Amplitude (mag)}			&
	\colhead{t$_3$ (days)}				&
	\colhead{$A-A_0$}					&
	\colhead{FWHM H$\alpha$}			&
	\colhead{He II}						&
	\colhead{Highest Iron}				&
	\colhead{Light Curve}				&
	\colhead{P$_\mathrm{orb}$ (days)}		&
	\colhead{$J - H$}					&		
	\colhead{$H - K$}					&	
	\colhead{Red Giant?}				&	
	\colhead{Other}					&
	\colhead{Category}					}	
\startdata	
LS And 	&	8.8	[1]$^a$	&	8	[2]	&	-1.64	&	...	   	&	...	   	&	...	   	&	...	   	&	...	   	&	...	   	&	...	   	&	...	   	&	Amp/$t_3$	[3]	&	F	\\
OS And 	&	11	[4]	&	23	[4]	&	2.63	&	1740	[5]	&	...	   	&	...	   	&	D	[4]	&	...	   	&	...	   	&	...	   	&	...	   	&	$M_\mathrm{WD}=1.0 M_\odot$	[6]	&	D	\\
DO Aql 	&	9.5	[4]	&	900	[7]	&	8.29	&	...	   	&	...	   	&	Fe III	[8]	&	F	[4]	&	0.168	[1]	&	0.554	[9]	&	0.277	[9]	&	No	[10]	&	...	   	&	E	\\
EL Aql 	&	13.6	[1]	&	25	[7]	&	5.39	&	...	   	&	...	   	&	...	   	&	...	   	&	...	   	&	0.95	[11]	&	0.17	[11]	&	Yes	[10]	&	...	   	&	D	\\
EY Aql 	&	9.5	[1]	&	...	   	&	...	&	...	   	&	...	   	&	...	   	&	...	   	&	...	   	&	...	   	&	...	   	&	...	   	&	...	   	&	F	\\
V356 Aql 	&	11.3	[4]	&	140	[4]	&	6.46	&	...	   	&	No	[8]	&	Fe II	[8]	&	J	[4]	&	...	   	&	0.85	[11]	&	-0.77	[11]	&	No	[10]	&	...	   	&	D	\\
V368 Aql 	&	10.4	[1]	&	42	[7]	&	3.20	&	...	   	&	Yes	[8]	&	Fe II	[8]	&	...	   	&	0.691	[12]	&	0.383	[9]	&	0.114	[9]	&	No	[10]	&	...	   	&	C	\\
V500 Aql 	&	11.2	[1]	&	43	[13]	&	4.06	&	...	   	&	...	   	&	...	   	&	...	   	&	0.145	[1]	&	...	   	&	...	   	&	...	   	&	...	   	&	F	\\
\enddata
\end{deluxetable}

\newpage
\begin{deluxetable}{l c c c c c c c c c c c c}
\rotate
\tablecaption{Properties of the Good Galactic Recurrent Nova Candidates}
\tabletypesize{\tiny}
\tablewidth{0pt}
\setlength{\tabcolsep}{0.02in}
\tablehead{
	\colhead{Nova}						&
	\colhead{Amplitude (mag)}			&
	\colhead{t$_3$ (days)}				&
	\colhead{Amp\slash t$_3$}			&
	\colhead{FWHM H$\alpha$}			&
	\colhead{He II}						&
	\colhead{Highest Iron}				&
	\colhead{Light Curve}				&
	\colhead{P$_\mathrm{orb}$ (days)}			&
	\colhead{$J - H$}					&		
	\colhead{$H - K$}					&	
	\colhead{Red Giant?}				&	
	\colhead{Other}					}	
\startdata	
V1721 Aql	&	$>$6	[1]	&	10:	[2]	&	...	&	6450	[2]	&	...	   	&	...	   	&	...	   	&	...	   	&	...	   	&	...	   	&	...	   	&	Triple-Peaked H$\alpha$	[3]	\\
DE Cir 	&	8.5	[4]	&	9.8	[1]	&	-1.54	&	5100	[5]	&	Yes	[6]	&	...	   	&	...	   	&	...	   	&	0.529	[7]	&	0.245	[7]	&	No	[8]	&	Triple-Peaked H$\alpha$	[6]	\\
CP Cru 	&	10.2	[4]	&	10	[9]	&	0.2	&	2000	[10]	&	Yes	[10]	&	...	   	&	...	   	&	0.944	[4]	&	...	   	&	...	   	&	...	   	&	...	   	\\
KT Eri	&	6.7	[1]	&	10	[11]	&	-3.3	&	3200-3400	[12]	&	Yes	[6]	&	No	[13]	&	P	[1]	&	737$^b$	[14]	&	0.46	[14]	&	0.07	[14]	&	Yes$^b$	[14]	&	Triple-Peaked H$\alpha$	[13]	\\
V838 Her 	&	13.8	[15]	&	4	[15]	&	2.01	&	5000	[16]	&	Yes	[17]	&	Fe II	[18]	&	P	[15]	&	0.297635	[4]	&	0.666	[7]	&	0.104	[7]	&	No	[8]	&	$M_\mathrm{WD}=1.35 M_\odot$, Triple-Peaked H$\alpha$	[19,20]	\\
V2672 Oph	&	$>$10	[1]	&	6:	[1]	&	...	&	8000	[21]	&	Yes	[22]	&	...	   	&	...	   	&	...	   	&	...	   	&	...	   	&	...	   	&	Triple-Peaked H$\alpha$	[22]	\\
V4160 Sgr 	&	$>$12	[15]	&	3	[15]	&	...	&	4500	[23]	&	Yes	[23]	&	...	   	&	S	[15]	&	...	   	&	...	   	&	...	   	&	...	   	&	...	   	\\
V4643 Sgr 	&	$>$8.3	[15]	&	6	[15]	&	...	&	4700	[24]	&	...	   	&	...	   	&	S	[15]	&	...	   	&	...	   	&	...	   	&	...	   	&	...	   	\\
\enddata
\tablenotetext{a}{Reference Key:
	1 = AAVSO database;
	2 = \citet{hounsell2011a};
	3 = \citet{helton2008b};
	4 = \citet{downes2001a};
	5 = \citet{liller2003a};
	6 = The STONY BROOK/SMARTS Spectral Atlas of Southern Novae;
	7 = Simbad database;
	8 = This Paper;
	9 = \citet{lyke2003a};
	10 = \citet{della-valle1997a};
	11 = \citet{hounsell2010a};
	12 = \citet{yamaoka2009a};
	13 = F. Walter 2010, private communication;
	14 = \citet{jurdana-sepic2012a};
	15 = \citet{strope2010a};
	16 = \citet{feast1991a};
	17 = \citet{iijima1991a};
	18 = \citet{della-valle1991a};
	19 = \citet{kato2009a};
	20 = \citet{iijima2010a};
	21 = \citet{ayani2009a};
	22 = \citet{schwarz2009a};
	23 = \citet{kingsburgh1991a};
	24 = \citet{ayani2001b};
	25 = \citet{vanlandingham2001a};
	26 = \citet{munari2006a};
	27 = \citet{fujii2005a};
	28 = \citet{mazuk2005a};
	29 = \citet{hachisu2007a};
}
\tablenotetext{b}{We note that the orbital period and red giant companion of KT Eri are disputed within the community, but all measures thus far put it easily in the ``Long orbital period" and ``Evolved companion of some sort" regime.}
\label{tab:candidates}
\end{deluxetable}

\newpage
\begin{deluxetable}{l c c}
\tablecaption{Recurrent Nova Category Fractions}
\tabletypesize{\scriptsize}
\tablewidth{0pt}
\tablehead{
	\colhead{ }				&
	\colhead{Complete Sample}	&
	\colhead{Strope Sample}		}	
\startdata	
A = \# known RNe	&	10	&	8	\\
B = \# strong candidate RNe	&	7	&	4	\\
C = \# likely RNe candidates	&	29	&	15	\\
D = \# likely CNe	&	91	&	45	\\
E = \# certain CNe	&	21	&	15	\\
F = \# not enough information	&	89	&	3	\\
A+B+C+D+E+F = sample size	&	247	&	90	\\
(B+C)/(B+C+D+E) = RN fraction	&	24.3\%$\pm$3.5\%	&	24.1\%$\pm$4.8\%	\\
\enddata
\label{tab:rnfraction}
\end{deluxetable}

\newpage
\begin{deluxetable}{l c c c c c c c}
\tablecaption{Recurrent Nova Discovery}
\tabletypesize{\scriptsize}
\tablewidth{0pt}
\tablehead{
	\colhead{Recurrent Nova}	&
	\colhead{$F_\mathrm{disc}$}		&
	\colhead{$\tau_\mathrm{rec}$}	&
	\colhead{$f_0$}			&
	\colhead{$f_1$}			&
	\colhead{$f_{\geq2}$}		&
	\colhead{$N_\mathrm{CN}$}			&
	\colhead{$N_\mathrm{RN}$}			}	
\startdata	
CI Aql	&	0.05	&	24	&	73\%	&	23\%	&	4\%	&	7 $\pm$ 2.2	&	28	\\
V394 CrA	&	0.013	&	30	&	93.4\%	&	6.4\%	&	0.2\%	&	31 $\pm$ 5.3	&	476	\\
T CrB$^a$	&	0.09	&	80	&	90\%	&	9\%	&	1\%	&	9 $\pm$ 2.9	&	100	\\
IM Nor	&	0.15	&	82	&	78\%	&	21\%	&	1.0\%	&	19 $\pm$ 3.9	&	91	\\
RS Oph	&	0.06	&	14.7	&	58\%	&	33\%	&	9\%	&	4 $\pm$ 1.6	&	11	\\
V2487 Oph$^b$	&	0.012	&	18	&	92.4\%	&	3.2\%	&	4.4\%	&	1 $\pm$ 1.0	&	23	\\
T Pyx$^b$	&	0.19	&	19	&	25\%	&	6\%	&	69\%	&	0 $\pm$ 0.3	&	2	\\
V3890 Sgr	&	0.03	&	25	&	88.3\%	&	11.2\%	&	0.5\%	&	23 $\pm$ 4.6	&	208	\\
U Sco$^b$	&	0.006	&	10.3	&	96.2\%	&	1.3\%	&	2.5\%	&	1 $\pm$ 0.7	&	40	\\
V745 Sco	&	0.014	&	21	&	92.7\%	&	7.1\%	&	0.2\%	&	44 $\pm$ 6.4	&	625	\\
\enddata
\tablenotetext{a}{For T CrB, the case is for undirected searches for both events with observations extended from 1850 to 1890 (with a 10\% detection efficiency).}
\tablenotetext{b}{The second-discovered eruption is taken to be a product of a directed nova search.}
\label{tab:rndiscovery}
\end{deluxetable}

\newpage
\begin{deluxetable}{l c c c c c c c c c c c c c c}
\rotate
\tablecaption{Archival Plate Search Status}
\tabletypesize{\tiny}
\tablewidth{0pt}
\setlength{\tabcolsep}{0.01in} 
\tablehead{
	\colhead{Plate Series$^a$}	&
	\colhead{V838 Her}			&
	\colhead{ }				&
	\colhead{V2487 Oph}		&
	\colhead{ }				&
	\colhead{V2672 Oph}		&
	\colhead{ }				&
	\colhead{V1172 Sgr}		&
	\colhead{ }				&
	\colhead{V4160 Sgr}		&		
	\colhead{ }				&	
	\colhead{V4643 Sgr}		&	
	\colhead{ }				&	
	\colhead{V723 Sco}			&
	\colhead{ }						
	}	
\startdata	
A$^b$	&	15$^c$	&		&	20	&		&	70	&		&	...	&		&	...	&		&	...	&		&	...	&		\\
AM/AC/AX/AY	&	200	&	(18$^h$, +15$^\circ$)	&	500	&	(17$^h$, -15$^\circ$)	&	500	&	(17$^h$, -15$^\circ$)	&	500	&	(17$^h$, -15$^\circ$)	&	150	&	(17$^h$, -15$^\circ$)	&	500	&	(17$^h$, -15$^\circ$)	&	...	&	(17$^h$, -30$^\circ$)	\\
AM/AC/AX/AY	&	700	&	(18$^h$, 0$^\circ$)	&	150	&	(17$^h$, -30$^\circ$)	&	150	&	(17$^h$, -30$^\circ$)	&	200	&	(18$^h$, -15$^\circ$)	&	...	&	(17$^h$, -30$^\circ$)	&	...	&	(17$^h$, -30$^\circ$)	&	300	&	(17$^h$, -45$^\circ$)	\\
AM/AC/AX/AY	&	700	&	(19$^h$, +15$^\circ$)	&	250	&	(18$^h$, -15$^\circ$)	&	300	&	(17$^h$, -45$^\circ$)	&	300	&	(18$^h$, -30$^\circ$)	&	300	&	(17$^h$, -45$^\circ$)	&	300	&	(17$^h$, -45$^\circ$)	&	300	&	(18$^h$, -30$^\circ$)	\\
AM/AC/AX/AY	&	300	&	(18$^h$, 0$^\circ$)	&	100	&	(18$^h$, -30$^\circ$)	&	200	&	(18$^h$, -15$^\circ$)	&	...	&		&	50	&	(18$^h$, -15$^\circ$)	&	200	&	(18$^h$, -15$^\circ$)	&	100	&	(18$^h$, -45$^\circ$)	\\
AM/AC/AX/AY	&	x$^d$	&		&	x	&		&	300	&	(18$^h$, -30$^\circ$)	&	x	&		&	300	&	(18$^h$, -30$^\circ$)	&	300	&	(18$^h$, -30$^\circ$)	&	x	&		\\
AM/AC/AX/AY	&	x	&		&	x	&		&	100	&	(18$^h$, -45$^\circ$)	&	x	&		&	100	&	(18$^h$, -45$^\circ$)	&	50	&	(18$^h$, -45$^\circ$)	&	x	&		\\
AI/BI	&	200	&	(19$^h$, +15$^\circ$)	&	...	&		&	200	&	(17$^h$, -15$^\circ$)	&	...	&		&	200	&	(17$^h$, -15$^\circ$)	&	200	&	(17$^h$, -15$^\circ$)	&	200	&	(17$^h$, -15$^\circ$)	\\
\enddata

\tablenotetext{a}{All plates are at the Harvard College Observatory, except for the last two series, which are at the Sonneberg Observatory.}
\tablenotetext{b}{The average time coverage for each series is as follows: A (1893-1950), AM/AC/AX/AY (1898-1953), AI/BI (1900-1953), B (1890-1953), Damons (1962-1989), I (1890-1946), IR (1934-1953), MC/MF/MA (1905-1953, 1960-1984), RH/RB (1928-1953), X (1890-1951), Sonneberg Patrols and Deep (1920-1945 sparse, 1945-present well covered)}
\tablenotetext{c}{The given numbers are good estimates, but not exact. On average they are correct within ${\sim}20\%$.}
\tablenotetext{d}{In this table, ``..." indicates that the series has not yet been checked, while ``x" indicates that the series does not have any coverage of the system.}

\label{tab:plates}
\end{deluxetable}

\end{document}